\documentclass[letterpaper,11pt]{article}
\usepackage[T1]{fontenc}
\usepackage[margin=3cm]{geometry}

\usepackage{amssymb,amsmath,amsthm,amsfonts}
\usepackage{mathtools} 

\usepackage{algorithm}
\usepackage{algorithmicx}
\usepackage[noend]{algpseudocode}
\usepackage{caption,subcaption}

\usepackage[table]{xcolor}
\usepackage{graphicx}

\usepackage{booktabs}
\usepackage{tabularx}
\usepackage{hyperref}

\usepackage[noabbrev,capitalize]{cleveref}
\usepackage{url}
\usepackage{xurl}

\usepackage{pgfplots}
\usepackage{pgfplotstable}
\pgfplotsset{compat=1.17}
\usepackage{wrapfig}

\usepackage{tikz}

\definecolor{yellow1}{rgb}{1,0.55,0} 
\definecolor{green1}{rgb}{0,0.4,0} 
\definecolor{blue1}{rgb}{0,0.1880,1}

\usepackage[round]{natbib}
\newcommand{\Tr}{\mathsf{T}}

\usepackage{hyphenat} 

\usepackage{nicefrac}       

\newcommand{\map}[2]{\text{map}_{#1\rightarrow#2}} 
\newcommand{\sss}[1]{\scriptscriptstyle{#1}} 
\newcommand{\varT}{\phi}
\newcommand{\varO}{\psi}

\usepackage[bb=dsserif]{mathalpha}

\begin{document}

\begin{center}
	{\huge\bf Marginalized particle Gibbs \\ for multiple state-space models \\ coupled through shared parameters \\}
\end{center}

\bigskip

\begin{minipage}{.45\textwidth}
	\begin{center}
		{\footnotesize{Anna Wigren}}\\
		{\scriptsize{Department of Information Technology,\\ Uppsala University, Sweden \\ anna.wigren@it.uu.se \\}}
	\end{center}
\end{minipage}%
\begin{minipage}{.5\textwidth}
	\begin{center}
		{\footnotesize{Fredrik Lindsten}} \\
		{\scriptsize{Department of Computer and Information Science, \\ Linköping University, Sweden \\ fredrik.lindsten@liu.se \\}}
	\end{center}
\end{minipage}

\bigskip

\begin{abstract}
	We consider Bayesian inference from multiple time series described by a common state-space model (SSM) structure, but where different subsets of parameters are shared between different submodels. An important example is disease-dynamics, where parameters can be either disease or location specific. Parameter inference in these models can be improved by systematically aggregating information from the different time series, most notably for short series. 
	Particle Gibbs (PG) samplers are an efficient class of algorithms for inference in SSMs, in particular when conjugacy can be exploited to marginalize out model parameters from the state update. 
	We present two different PG samplers that marginalize static model parameters on-the-fly: one that updates one model at a time conditioned on the datasets for the other models, and one that concurrently updates all models by stacking them into a high-dimensional SSM. 
	The distinctive features of each sampler make them suitable for different modeling contexts. We provide insights on when each sampler should be used and show that they can be combined to form an efficient PG sampler for a model with strong dependencies between states and parameters.
	The performance is illustrated on two linear-Gaussian examples and on a real-world example on the spread of mosquito-borne diseases.
\end{abstract}

\section{Introduction} \label{sec:Intro}
The nonlinear state-space model (SSM), which relates latent states with observations from some process, is a commonly used model structure in many areas of application such as control \citep{Astrom2012}, epidemiology \citep{Funk2016}, ecology \citep{Finke2019} and climatology \citep{Calafat2018}. Sometimes, there may be several datasets available that stem from different dynamical processes, but that share some traits (parameters) despite being otherwise distinct from each other. One example is datasets from outbreaks of different mosquito-borne diseases, where either location-specific parameters (e.g.\ mosquito density) or disease-specific parameters (e.g.\ incubation time) are shared between different outbreaks. Intuitively, it should be beneficial to use information from all datasets when identifying the parameters, in particular if the datasets are small. For this purpose, the SSMs describing each outbreak can be combined into a \textit{multi-SSM} structure, where the shared parameters connect the outbreak-specific models.
In this paper we present methods for efficient Bayesian inference in multi-SSMs that are coupled through shared parameters.

Bayesian inference in SSMs is in general intractable and approximate methods, such as Markov chain Monte Carlo (MCMC), must be used \citep{Robert2004}. The focus here is on the Gibbs sampler \citep{Geman1984}, one particular instance of MCMC that draws samples from the desired posterior distribution by alternating between sampling new state trajectories and sampling new parameters. Generating new state trajectories can be problematic due to them being both high-dimensional and highly auto-correlated. Fortunately, a Markov kernel based on sequential Monte Carlo (SMC) \citep{Chopin2020,Naesseth2019} can be applied for approximating the state update. The resulting method, particle Gibbs (PG), is an ``exact approximation'' of the Gibbs sampler in the sense that it has the correct stationary distribution for any number of particles, and as the number of particles used in the Markov kernel increases, the PG sampler approaches the ideal Gibbs sampler in terms of autocorrelation \citep{Andrieu2010}. It can, however, never surpass the Gibbs sampler. The same is true for extensions of PG, such as the more efficient particle Gibbs with ancestor sampling (PGAS) \citep{Lindsten2014}. This is problematic, since the Gibbs sampler (and thereby also PG and PGAS) is known to produce highly correlated samples when there is a strong dependence between the state trajectory and the parameters. To address this issue, it was suggested by \cite{Wigren2019} to marginalize out the parameters from the state update in the PG sampler. The marginalized PG sampler approximates a sampler that, if it could be realized, would generate independent samples from the target distribution. In this article we show how to extend this approach to the multi-SSM setting. 

It turns out that there are several ways to construct marginalized PG samplers for multi-SSMs; we develop two distinct techniques.
The first sampler, \textit{single marginalized PG}, updates each individual SSM one at a time, conditioned on the current state trajectories for the other datasets that it shares parameters with. We show that this sampler can be formulated in a modular fashion, where the outer module handles the conditioning on other datasets and the inner module is equivalent to the sampler by \cite{Wigren2019}. The modularity makes the sampler well-suited for implementation in probabilistic programming languages (PPLs) \citep{Meent2018,Murray2018a}. Furthermore, by updating with respect to one dataset at a time the dimension of the SMC-based Markov kernels can be kept low even when the number of datasets increase. This can have implications in practice, since SMC is known to suffer from weight degeneracy for high-dimensional problems \citep{Bickel2008,Snyder2008,Snyder2015}. The main drawback of this sampler is that the conditioning on the other datasets introduces extra correlation between the samples, implying that it is unable to generate independent and identically distributed (iid) samples even when the number of particles approaches infinity.

To address the potential shortcomings of the first sampler, we also consider a sampler with complementary attributes. Rather than updating each dataset individually, the \textit{stacked marginalized PG} sampler instead treats the multi-SSM as one joint SSM by stacking the datasets together into one higher-dimensional SSM and then marginalizes this new model with respect to the parameters. The stacking of datasets can be either with respect to the time or the state dimension. We show that for both stacking techniques the marginalized sampler resembles the method by \cite{Wigren2019}, but the stacking of datasets introduces dependencies between different parts of the SSM that must be incorporated into the base algorithm. 
Provided that enough particles are used in the SMC-based Markov kernel, the stacked marginalized PG sampler can yield (close to) iid samples from the target distribution. However, if the number of datasets is large the higher-dimensional SSM can quickly cause the SMC-based Markov kernel to degenerate, unless a very large number of particles is used. 
The stacked sampler is most useful for models with strong depedencies between states and observations since it, in contrast to the single marginalized PG sampler, does not use any information from the other datasets during the first iterations in the SMC-based kernel. 

In essence, the two marginalized PG samplers represent two extremes. 
However, as both samplers are valid MCMC methods, it is possible to combine them to obtain a method that attains a mix of their properties. We consider one such combination aimed to loosen strong dependencies between individual updates for a mosquito-borne disease multi-SSM by stacking subsets of the SSMs and then do individual updates of the stacked subsets.

Parallels can be drawn between the design of marginalized PG samplers for multi-SSMs and the blocked PG schemes described by \cite{Murphy2016}, in the sense that both consider PG style samplers for high-dimensional SSMs with interactions between the different components. While \cite{Murphy2016} consider blocking of interacting systems in the general case, we study the particular instance of interacting systems stemming from marginalization of the system parameters and potential performance gains related to how this marginalization is designed. Nevertheless, the conclusions presented by \cite{Murphy2016} on the existence of a problem dependent optimal block size, and the interpretation that the blocking scheme represents a trade-off between efficiency gains from joint updates of correlated states and degeneracy problems in the SMC-based Markov kernel applies also to the design of marginalized PG samplers for multi-SSMs. 

Analogously to \cite{Wigren2019}, marginalization yields a non-Markovian model with dependencies between the datasets. In the general case, a non-Markovian model results in a quadratic computational complexity in the number of time-steps, which can be prohibitive in practice for longer time series. However, when there is conjugacy between the parameter prior and the complete data likelihood for each dataset, the required updates in each iteration can be done using sufficient statistics, which keeps the computational cost linear with respect to the lengths of the datasets.

Improving the performance of joint state and parameter inference using marginalization has been considered before, although not in the context of multi-SSMs. Apart from the work on the marginalized PG sampler for one dataset by \cite{Wigren2019}, marginalization of static parameters is also considered by \cite{Carvalho2010,Storvik2002}. There, the parameter posterior is assumed to be related to the latent states only through sufficient statistics that are incorporated into the state and inferred together with the states using SMC. Whether these methods are affected by path degeneracy can be discussed, see e.g.\ \citep{Chopin2010}. The marginalized PG samplers that we present here are also affected by path degeneracy, but not to the same extent since they belong to the particle MCMC \citep{Andrieu2010} family of methods. If path degeneracy is an issue, the more robust version of PG, PGAS \citep{Lindsten2014}, can be employed. The extension is straightforward (see \cite{Wigren2019} for details) and will likely result in a performance gain, but we do not elaborate on this possibility for brevity.

The article is structured in the following way: Section \ref{sec:notation} introduces the notation used in the article and formally defines the multi-SSM, whereas Section \ref{sec:PGmulti} provides a background on SMC and defines the PG sampler that will be used as a baseline in all comparisons. In Section \ref{sec:multiSSM_mPG} the different marginalized PG samplers we consider are introduced and their different characteristics are discussed. This section also includes experiments illustrating the performance of the marginalized PG samplers in different settings for simulated data.
Section \ref{sec:VBD} contains a simulation study with real-world data from outbreaks of mosquito-borne diseases and illustrates how the marginalized PG samplers introduced in Section \ref{sec:multiSSM_mPG} can be combined. Section \ref{sec:Discussion} contains a discussion of the main results. 
A derivation of main results, detailed descriptions of the experiments and additional results are available in the supplementary material.

\section{Model Structure and Notation} \label{sec:notation}
In a SSM, latent states $x_t$ evolve over time according to a Markovian transition density $p(x_t | x_{t-1},\theta)$, where $\theta$ are the parameters of the model. The initial state is drawn from an initial density $p(x_0 | \theta)$. Observations from the system are independent given the current state $x_t$, with a density $p(y_t | x_t, \theta)$. The complete collection of states, the state trajectory, is denoted $x_{0:T} = \{x_0,\dots,x_T\}$, and the sequence of observations is $y_{1:T} = \{y_1,\dots,y_T\}$. In the interest of space and readability, the short-hand notation $X=x_{0:T}$ and $Y=y_{1:T}$ is used where possible.

\begin{figure}
	\begin{center}
		\includegraphics[width=.3\textwidth]{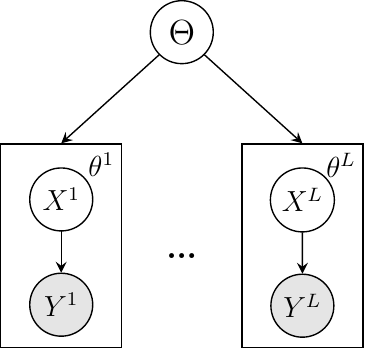}
		\caption{A general multi-SSM. There are $L$ datasets of observations $\{Y^\ell\}_{\ell=1}^L$, each with its own SSM (black rectangles) with parameters $\theta^\ell$. The parameters $\theta^\ell$ for dataset $\ell$ is a subset of the complete parameter vector $\Theta$ and one or more of these parameters are shared with one or more of the other datasets. 
		}
		\label{fig:hierSSMgen}	
	\end{center}
\end{figure}

A \textit{multi-SSM} can be constructed when $L$ datasets of observations $\{Y^\ell\}_{\ell=1}^L$, each of length $T^\ell$, are available and every dataset $Y^\ell$ is described by an individual SSM that shares one or more parameters with the other SSMs. 
More formally, all $K$ unique parameters of the model are collected in the joint parameter vector $\Theta\in\mathbb{R}^K$. 
We write $I^\ell \subset \{1,\dots, K\}$ for the index set indicating which parameters that affect submodel $\ell$, and $\theta^\ell$ for the corresponding subvector of parameters of dimension $K^\ell = | I^\ell |$. 
The typical situation that we envision is that the multi-SSM structure does not consist of separate, disconnected multi-SSMs, and for each dataset $\ell$, $I^\ell \cap I^j \neq \emptyset$ for at least one $j\neq\ell$. Note that this includes the case $I^\ell=I^j$, even when it holds for all $j$.
The multi-SSM is depicted in Figure \ref{fig:hierSSMgen}. Additionally, the complete data collection $\{ X^\ell, Y^\ell \}_{\ell=1}^L$ is denoted by $\mathcal{D}$, and $\mathcal{D}^{\neg \ell} = \{ X^j,Y^j \}_{j\neq \ell}$ denotes all state and observation sequences except $\ell$. 

To enable mapping between parameter vectors or sufficient statistics (see Section \ref{sec:multiSSM_mPG} for details) for different submodels we introduce three operators. First, $V=\text{expand}_\ell(v)$ takes a vector $v$ of dimension $K^\ell$ and returns a vector $V$ of dimension $K$, where the $k$th element is given by the corresponding value of $v$ if $k \in I^\ell$, and is zero otherwise. Second, $v = \text{reduce}_\ell(V)$ takes a vector of dimension $K$ and returns the subvector of dimension $K^\ell$ obtained by picking out the elements indexed by $I^\ell$. Finally, for notational brevity, we define $v^\ell = \map{j}{\ell}(v^j) = \text{reduce}_\ell(\text{expand}_j(v^j))$ as the composition of the two mappings. Note that the $\map{j}{\ell}$ operator produces a vector $v^\ell$ by copying elements from $v^j$ for parameter indices shared by submodels $j$ and $\ell$, and which is zero for indices not shared by the two models.

\section{Particle Gibbs for Multi-SSMs} \label{sec:PGmulti}
PG samplers use SMC to approximate the state update. In this section, we therefore first give an introduction to SMC before discussing how a PG sampler can be formulated for the multi-SSM. The topics in this section provide a background for the discussion on marginalized PG samplers for multi-SSMs, so we keep the discussion brief and refer readers to \cite{Chopin2020,Naesseth2019} for a detailed description of SMC and an introduction to PG.

\subsection{Sequential Monte Carlo} 
SMC uses a set of $N$ weighted samples (particles), denoted $\{ x_{0:t}^i, \bar{w}_t^i \}_{i=1}^N$, to approximate a sequence of target densities $\bar{\gamma}_{\theta,t}(x_{0:t})=\frac{\gamma_{\theta,t}(x_{0:t})}{Z_{\theta, t}}, \hspace{2mm} t=1,2,\dots,$ where $\gamma_{\theta,t}(x_{0:t})$ is the unnormalized target density and $Z_{\theta, t}$ is a normalization constant. For the case of a SSM, $\bar{\gamma}_{\theta,t}(x_{0:t})=p(x_{0:t} | y_{1:t},\theta)$ is a common target with $\gamma_{\theta,t}(x_{0:t})=p(x_{0:t}, y_{1:t} | \theta)$ and $Z_{\theta, t} = p(y_{1:t} | \theta)$.
The weighted samples are generated in an iterative process in three steps. First, resampling is performed by selecting ancestor trajectories $x_{0:t-1}^{a_t^i}$ according to the normalized weights $\bar{w}_{t-1}^i$ from the previous iteration. Next, the resampled trajectories are propagated to the next time using a proposal distribution $q_{\theta,t}(x_t | x_{0:t-1})$ chosen by the user. The transition density $p(x_t | x_{t-1}, \theta)$ is a popular choice due to its simplicity, but other more sophisticated proposal distributions are also commonly used, see e.g.\ \cite{Doucet2000}. Lastly, new weights are computed for the propagated trajectories according to the weight function 
\begin{align} \label{eq:SMCweight}
\omega_{\theta,t}(x_{0:t}) = \frac{\gamma_{\theta,t}(x_{0:t})}{\gamma_{\theta,t-1}(x_{0:t-1})q_{\theta,t}(x_{t} | x_{0:t-1})}.
\end{align}
The procedure is summarized in Algorithm \ref{alg:multiSSM_SMC}.

While SMC is well-suited for inference in non-linear models with a sequential structure, it is also prone to \textit{weight degeneracy}, in particular when the state $x_t$ is high-dimensional. It has been shown that to avoid weight degeneracy the number of particles must increase exponentially with the state dimension \citep{Bickel2008, Snyder2008,Snyder2015}. This can be prohibitive computation-wise and constitutes an important factor when designing a marginalized PG sampler for multi-SSMs.

\begin{algorithm} 
	\caption{Sequential Monte Carlo (all steps for $i=1,\dots,N$)}
	\label{alg:multiSSM_SMC}
	\begin{algorithmic}
		\State \textbf{Initialize:} Draw $x_{0}^{i} \sim q_{\theta,0}(x_{0})$ independently. Set $\bar{w}_0^ i = 1/N$. 
		\For {$t=1 \dots T$}
		\State (a) \textit{Resample:}
		Draw ancestor indices
		$a_t^i \sim \text{Categorical}(\{\bar{w}_{t-1}^ i\}_{i=1}^N)$.
		Set $\bar{w}_ {t-1}^ i=1/N$.
		\State (b) \textit{Propagate:} Simulate $x_ t^{i} \sim q_{\theta,t}(x_ t|x_{0:t-1}^{a_t^i})$. Set $x_{0:t}^{i}=\{ x_{0:t-1}^{a_t^i}, x_{t}^{i} \}$.
		\State (c) \textit{Weight:} Set $w_{t}^{i}=\omega_{\theta,t}(x_{0:t}^{i})$ and normalize $\bar{w}_ t^i=w_ t^i/ \sum\limits_{j=1}^N w_ t^j$. 
		\EndFor
	\end{algorithmic}
\end{algorithm}

\subsection{Particle Gibbs} \label{sec:PG}
In the multi-SSM (Figure \ref{fig:hierSSMgen}) all state and observation trajectory pairs $\{ X^\ell, Y^\ell \}_{\ell=1}^L$ are independent given the parameters. Thus, we can construct a PG sampler that first samples new parameter values and then updates the state trajectories one at a time. After initialization at $m=1$, such a PG sampler iterates, for $m = 2,\, 3,\, \dots$
\begin{align} \label{eq:PGforHSSM}
&\Theta(m) \sim p\big(\Theta | \mathcal{D}(m-1)\big) \nonumber \\
&\text{for }\ell=1\dots L  \\
&\hspace{7mm} X^{\ell}(m) \sim \kappa^{\theta^\ell(m)} \big(X^{\ell}(m-1),X^{\ell}\big), \nonumber
\end{align}
where $\kappa^{\theta^\ell}$ is a SMC-based Markov kernel, with $p\big(X^{\ell} | Y^{\ell},\theta^\ell \big)$ as stationary distribution, that stochastically maps a reference trajectory $X^{\ell}(m-1)$ from the previous iteration to a new trajectory $X^\ell(m)$. The Markov kernel applies conditional SMC (cSMC) to generate the new state trajectory. The key difference from the standard SMC procedure in Algorithm~\ref{alg:multiSSM_SMC} is that the reference trajectory is guaranteed to survive all resampling steps. This is achieved by setting $a_t^N=N$ in step (a), and $x_t^N=\tilde{x}_t$ in step (b) of the algorithm, where ``tilde'' is used to indicate the reference trajectory (that is, at iteration $m$ we have $\tilde x_t = x_t(m-1)$). As a final step we draw a new reference trajectory based on the weights at timestep $T$. Updating each state trajectory individually, like in \eqref{eq:PGforHSSM}, rather than jointly is beneficial when SMC is used, since degeneracy of the particle weights poses less of a problem for lower-dimensional updates. However, there is often a strong dependence between the states and parameters in SSMs, resulting in that the sampler produces highly correlated samples.

\section{Marginalized Particle Gibbs for Multi-SSMs} \label{sec:multiSSM_mPG}
The PG sampler in \eqref{eq:PGforHSSM} is often slow-mixing due to strong dependencies between the parameters and states. Inspired by the promising results by \cite{Wigren2019} for a single SSM, we wish to reduce the correlation of the PG sampler for multi-SSMs by eliminating the parameters from the state update. There are, in fact, several ways to design a marginalized PG sampler for a multi-SSM. We consider two different strategies in this section and illustrate their properties and performance on two examples with simulated data. 

The first strategy, derived and described in Section \ref{sec:single}, directly marginalizes the PG sampler in \eqref{eq:PGforHSSM}. After marginalization of the parameters, this sampler still updates one dataset at a time, but the update now depends on the other datasets that share the marginalized parameters. This sampler is referred to as \textit{single marginalized PG}. The second strategy, described in Section \ref{sec:stack}, instead stacks the datasets --- either with respect to the time or state dimension --- into one high-dimensional SSM. The parameters are eliminated from the new model and the marginalized sampler updates all datasets simultaneously. This sampler is referred to as \textit{stacked marginalized PG}.

\subsection{Single Marginalized Particle Gibbs for Multi-SSMs} \label{sec:single}
The first approach we consider is to use a marginalized SMC-based kernel that updates each trajectory individually, i.e.\, a marginalized version of \eqref{eq:PGforHSSM}. 
After initializing the sampler at $m=1$, it iterates, for $m = 2,\, 3,\, \dots$,
\begin{align} \label{eq:mPGforHSSM}
&\text{for }\ell=1\dots L \nonumber \\
&\hspace{7mm} X^{\ell}(m) \sim \kappa^\ell \Big(\{X^{1:\ell-1}(m),X^{\ell:L}(m-1)\},X^{\ell}\Big),
\end{align}
where $\kappa^\ell$ is a marginalized SMC-based Markov kernel with $p \big(X^{\ell} | Y^{\ell}, \mathcal{D}^{\neg \ell} \big)$ as its stationary distribution.
The parameters have been excluded from the procedure to emphasize that they have been \textit{eliminated} in the marginalized framework. If they are required, they can either be sampled in each iteration given the current state trajectories, or, if the state trajectories are stored, they can be generated after the final Gibbs iteration.
One advantage of this marginalization approach is that the individual updates of each state trajectory implies that the dimension is kept low(er) in the Markov kernels. This is beneficial, since the kernels are SMC-based, hence a too high dimension (in state or time) can cause weight or path degeneracy problems that renders the cSMC updates inefficient \citep{Bickel2008,Lindsten2015,Snyder2008,Snyder2015}. 
On the other hand, updating one trajectory at a time conditioned on all other trajectories will not, in contrast to the sampler by \cite{Wigren2019}, generate iid samples from the target distribution in the limit $N\to\infty$. Indeed, we obtain samples that are correlated, but less so than if the PG sampler in \eqref{eq:PGforHSSM} is used.

To fully specify the marginalized PG sampler in \eqref{eq:mPGforHSSM}, the SMC-based Markov kernel that generates the state trajectories must be derived. 
In the work by \cite{Wigren2019}, the marginalization of parameters resulted in a non-Markovian state update that could only be evaluated exactly for models exhibiting conjugacy between the parameter prior and the complete data likelihood. This holds true also for the multi-SSM, but with the addition that when the parameters are marginalized out from the update of one dataset, a dependence on the remaining datasets that share one or more of the parameters is introduced.

More formally, the marginalized target density at time $t$ of the cSMC kernel in \eqref{eq:mPGforHSSM} is $ \bar{\gamma}_t(x_{0:t}^\ell) = p\big(x_{0:t}^{\ell} | y_{1:t}^{\ell}, \mathcal{D}^{\neg \ell} \big)$, implying that to compute, e.g.\, the weights \eqref{eq:SMCweight} we must evaluate integrals of the form
\begin{align} \label{eq:margInt}
\int p(x_t^\ell,y_t^\ell | x_{t-1}^\ell,\theta^\ell)p(\theta^\ell | x_{0:t-1}^\ell,y_{1:t-1}^\ell,\mathcal{D}^{\neg \ell})\mathrm{d}\theta^\ell,
\end{align}
where $p(\theta^\ell | x_{0:t-1}^\ell,y_{1:t-1}^\ell,\mathcal{D}^{\neg \ell})$ is the posterior distribution of the parameters in the preceding timestep.  
In general, integrals like \eqref{eq:margInt} are not tractable unless there is a conjugacy relation between the parameter posterior and the joint observation and state likelihood. We therefore consider multi-SSMs where both distributions belong to the exponential family. 
The joint observation and state likelihood can thus be written as
\begin{align} \label{eq:EfamLhood}
p(x_t^\ell,y_t^\ell | x_{t-1}^\ell,\theta^\ell) = h_t^\ell \exp \big( {\theta^{\ell}}^\Tr s_t^\ell-{A^\ell}^\Tr (\theta^\ell)r_t^\ell \big),
\end{align}
where $h_t^\ell = h^\ell(x_t^\ell,x_{t-1}^\ell,y_t^\ell)$ is the base measure, $s_t^\ell = s^\ell(x_t^\ell,x_{t-1}^\ell,y_t^\ell)$ is a sufficient statistic and $A^\ell(\theta^\ell)$ is the log-partition function. The factor $r_t^\ell$ can be viewed as an additional statistic connected to the hyperparameter $\nu_0^\ell$ in the conjugate prior. It is introduced to facilitate the mapping of statistics described later in this section\footnote{Apart from facilitating the mapping of statistics, this factor can also incorporate dependencies on previous states in the log-partition function that would otherwise prevent formulating a conjugate prior for the complete data likelihood. A detailed description is available in the supplementary material of \cite{Wigren2019}.} and will in all cases considered in this article be a vector of ones, i.e.\ $r_t^\ell=\mathbb{1}^{\sss{K^\ell}}$.
The conjugate prior to \eqref{eq:EfamLhood} is 
\begin{align} \label{eq:EfamPrior}
p(\theta^\ell | \chi^\ell_0, \nu^\ell_0) = g^\ell(\chi^\ell_0,\nu^\ell_0) \exp \big( {\theta^{\ell}}^\Tr \chi^\ell_0 - {A^\ell}^\Tr (\theta^\ell)\nu^\ell_0 \big),
\end{align}
where $\chi^\ell_0$ and $\nu^\ell_0$ are hyperparameters and $g^\ell$ is a normalizing factor. 

For a conjugate model, the parameter posterior in \eqref{eq:margInt} belongs to the same parametric family of distributions as the prior \eqref{eq:EfamPrior}, but its hyperparameters have been updated with data from earlier timesteps. 
Therefore, when sampling dataset $\ell$ in the multi-SSM, we can carry out the marginalization analogously to the single dataset case, i.e.\ by simply replacing the hyperparameters of the prior distribution with those of the posterior and evaluate \eqref{eq:margInt} analytically.
In fact, all expressions required for the marginalized cSMC kernel in \eqref{eq:mPGforHSSM} are completely analogous to those derived by \cite{Wigren2019} apart from the hyperparameter updates, which include an extra term to account for the posterior's dependence on the other datasets. 
The parameter posterior is proportional to the product of the prior and the complete likelihood (up until the current timestep). Consequently, the updated hyperparameters at timestep $t$ will consist of sums of sufficient statistics, both from dataset $\ell$ and from the remaining datsets that we condition the $\ell$th update on. 
The hyperparameters at timestep $t$ are
\begin{align} 
&\chi_{t}^\ell = \chi^\ell_0 + \sum_{j\neq\ell} \map{j}{\ell}(S^j) + \sum_{k=1}^{t}s_k^\ell = \bar{\chi}_0^\ell + \sum_{k=1}^{t}s_k^\ell, \label{eq:hyperParam1} \\
&\nu_{t}^\ell = \nu^\ell_0 + \sum_{j\neq\ell} \map{j}{\ell}(R^j) + \sum_{k=1}^{t}r_k^\ell = \bar{\nu}_0^\ell + t\cdot\mathbb{1}^{\sss{K^\ell}}, \label{eq:hyperParam2}
\end{align}
with $S^j=\sum\limits_{t=1}^{T_j}s_t^j$ and $R^j=T^j\cdot\mathbb{1}^{\sss{K^j}}$ being sums of statistics. 
The operator $\map{j}{\ell}$ was defined in Section \ref{sec:notation} and ensures that only statistics from datasets that share parameters with dataset $\ell$ are added in the hyperparameter updates. 
The extra term in \eqref{eq:hyperParam1}-\eqref{eq:hyperParam2} compared to the single dataset case is a sum over the statistics for the datasets that we condition on. 
All terms in the middle sum are independent of $t$, i.e.\ fixed during the update of dataset $\ell$, and can be included in the initial hyperparameters, now denoted $\bar{\chi}_0^\ell$ and $\bar{\nu}_0^\ell$ (second equality in \eqref{eq:hyperParam1}-\eqref{eq:hyperParam2}). The contribution from the other datasets to the initial hyperparameters has the same effect as an informative prior on the parameters would have. This implies that the blocking strategy that was suggested by \cite{Wigren2019} to mitigate the effect of poor mixing during the initial timesteps for uninformative priors is often not required for the sampler in \eqref{eq:mPGforHSSM}
A complete derivation of all expressions required for the cSMC kernel is given in Section \ref{app:single} of the supplementary material.

Including the dependence on the other datasets in the initial hyperparameters enables formulating the algorithm in a modular fashion, where the inner module is exactly analogous to the marginalized cSMC kernel by \cite{Wigren2019} operating on one dataset at a time. The outer module is the complete single marginalized PG sampler for multi-SSMs. It iterates over all datasets by first updating the initial hyperparameters for dataset $\ell$ using the sufficient statistics from the other datasets, and then updates the current state trajectory by applying the inner module. The single marginalized PG sampler is given in Algorithm~\ref{alg:multiSSM_mPG} and the marginalized cSMC kernel is given in Algorithm~\ref{alg:multiSSM_mcSMC}. The modular structure makes it easy to replace parts of the algorithm, e.g.\ we get a marginalized PGAS sampler \citep{Lindsten2014} for multi-SSMs by simply replacing the inner cSMC kernel with one that applies marginalized ancestor sampling \citep{Wigren2019}. In particular, the modular structure has important implications for the usability of single marginalized PG in PPLs. If a framework for encoding the multi-SSM structure and initial hyperparameters (outer module) is provided in the PPL, it suffices to implement Algorithm~\ref{alg:multiSSM_mcSMC} to incorporate automatic single marginalized PG for multi-SSM structures.

\begin{algorithm} 
	\caption{Single marginalized particle Gibbs for multi-SSMs.}
	\label{alg:multiSSM_mPG}
	\begin{algorithmic}
		\State \textbf{Input:} Observations $\{Y^\ell\}_{\ell = 1}^L$, initial hyperparameters $\chi_0$, $\nu_0$. 	
		\State \textbf{Initialize:} Generate trajectories $\{X^\ell(1)\}$ from Algorithm \ref{alg:multiSSM_SMC} for some initial $\theta$, compute their sums of statistics $\{S^\ell, R^\ell\}_{\ell=1}^L$.
		\For {$m=2 \dots M$}
		\For {$\ell=1 \dots L$}
		\State (a) Update hyperparameters $\bar{\chi}^{\ell}_0=\chi^\ell_0 + \sum\limits_{j\neq\ell}\map{j}{\ell}(S^j)$,  $\bar{\nu}^{\ell}_0=\nu^\ell_0+\sum\limits_{j\neq\ell} \map{j}{\ell}(R^j)$. 
		\State (b) Run Algorithm \ref{alg:multiSSM_mcSMC} with inputs $X^\ell(m-1)$, $\bar{\chi}^{\ell}_0$, $\bar{\nu}^{\ell}_0$, $Y^\ell$.
		\State (c) Store the output from Algorithm \ref{alg:multiSSM_mcSMC} as $X^\ell(m) = \tilde{X}$, $S^\ell=\tilde{S}$ and  $R^\ell=\tilde{R}$.
		\EndFor
		\EndFor
		\State \textbf{Output:} State trajectories $\{ \{X^\ell(m)\}_{m=1}^M \}_{\ell=1}^L$. 
	\end{algorithmic}
\end{algorithm}

\begin{algorithm} 
	\caption{Marginalized conditional SMC kernel (all steps for $i=1,\dots,N$)}
	\label{alg:multiSSM_mcSMC}
	\begin{algorithmic}
		\State \textbf{Input:} Proposal $q_{t}$, base measure $h_t$, normalizing factor $g$, initial hyperparameters $\bar{\chi}$ and $\bar{\nu}$, reference trajectory $\tilde{x}_{0:T}$, observations $y_{1:T}$.		
		\State \textbf{Initialize:} Draw $x_{0}^{i} \sim q_{0}(x_{0})$ independently, set $x_{0}^{N}=\tilde{x}_{0}$, $S_0^i=0$, $R_0^i=0$. Set $\bar{w}_0^ i = 1/N$. 
		\For {$t=1 \dots T$}
		\State (a) \textit{Resample:} Draw 
		$a_t^i \sim \text{Categorical}(\{\bar{w}_{t-1}^ i\}_{i=1}^N)$
		and set $a_ t^N=N$. Set  $\bar{w}_ {t-1}^ i=1/N$.
		\State (b) \textit{Propagate:} Simulate $x_ t^{i} \sim q_{t}(x_ t|x_{0:t-1}^{a_t^i})$ and set $x_{t}^{N}= \tilde{x}_{t}$.  Set $x_{0:t}^{i}=\{ x_{0:t-1}^{a_t^i}, x_{t}^{i} \}$.
		\State (c) \textit{Update sum of statistics:} $S_t^i=S_{t-1}^i+s_t^i$, $R_t^i=R_{t-1}^i+1$.
		\State (d) \textit{Weight:} Set $w_{t}^{i}=h_t^i \frac{g^\ell(\bar{\chi}+S_{t-1}^i, \bar{\nu}+R_{t-1}^i)}{g^\ell(\bar{\chi}+S_{t}^i, \bar{\nu}+R_{t}^i)q_{t}(x_t^i|x_{t-1}^{a_t^i})}$ and  normalize $\bar{w}_ t^ i=w_ t^i / \sum\limits_{j=1}^N w_ t^ j$.
		\EndFor
		\State \textbf{Output:} Draw 
		$k \sim \text{Categorical}(\{\bar{w}_{T}^{ i}\}_{i=1}^{N})$ and
		output new reference trajectory $\tilde{x}_{0:T}=x_{0:T}^{k}$ and its sum of statistics $\tilde{S} = S_{T}^k$, $\tilde{R} = R_{T}^k$. 
	\end{algorithmic}
\end{algorithm}

\subsubsection{Illustration: Single Marginalized Particle Gibbs for a Linear\hyp{}Gaussian Multi-SSM} \label{sec:LGhier}
\begin{figure}
	\begin{center}
		\includegraphics[width=.35\textwidth]{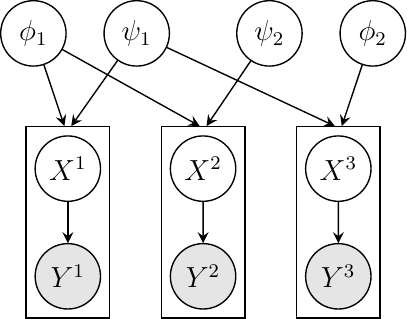}
		\caption{A linear-Gaussian multi-SSM for the case of three datasets and four parameters: the process noise variances $\varT_1$ and $\varT_2$, and the measurement noise variances $\varO_1$ and $\varO_2$. Dataset 1 and 2 share process noise variance and dataset 1 and 3 share observation noise variance.}
		\label{fig:LGhierSSM}
	\end{center}
\end{figure}

As a first example we consider a linear-Gaussian multi-SSM with three datasets of different lengths and four parameters. Each individual SSM has the following transition and observation densities
\begin{align} \label{eq:LGSSM1}
p(x_t| x_{t-1},\theta) &= \mathcal{N}(x_t | ax_{t-1},\varT), \\
p(y_t | x_t,\theta) &= \mathcal{N}(y_t | cx_t,\varO), \label{eq:LGSSM2}
\end{align}
where $a=0.9$, $c=1$ and $\mathcal{N}(z| \mu, \sigma^2)$ denotes the Gaussian distribution of a random variable $z$ with mean $\mu$ and variance $\sigma^2$. The noise variances $\varT$ and $\varO$ are unknown and shared between datasets in the multi-SSM. For the purpose of this example, dataset~1 and~2 share transition model (parameter $\varT_1$) and dataset~1 and~3 share observation model (parameter $\varO_1$). This multi-SSM is depicted in Figure \ref{fig:LGhierSSM}.

First, we illustrate the benefit of using the single marginalized PG sampler (Algorithm \ref{alg:multiSSM_mPG}) as opposed to the standard PG sampler in \eqref{eq:PGforHSSM}. We simulated $T^1=30$, $T^2=50$ and $T^3=40$ observations from SSM 1, 2 and 3, respectively. To allow for marginalization, we used inverse gamma priors on the noise variances. Both algorithms were run for $M=20000$ iterations and the first 2000 samples were discarded as burn-in. The standard PG sampler will, by construction, generate one sample from the current parameter posterior in each iteration. The marginalized PG sampler (Algorithm \ref{alg:multiSSM_mPG}) only generates state trajectory samples, but we have added a step that samples from the parameter posterior in each iteration to enable comparing the quality of parameter samples from the two algorithms. Additional details about the experiments are available in Section \ref{app:impDetailLG}.

\begin{figure}[ht]
	\centering
	\includegraphics[width=.8\textwidth]{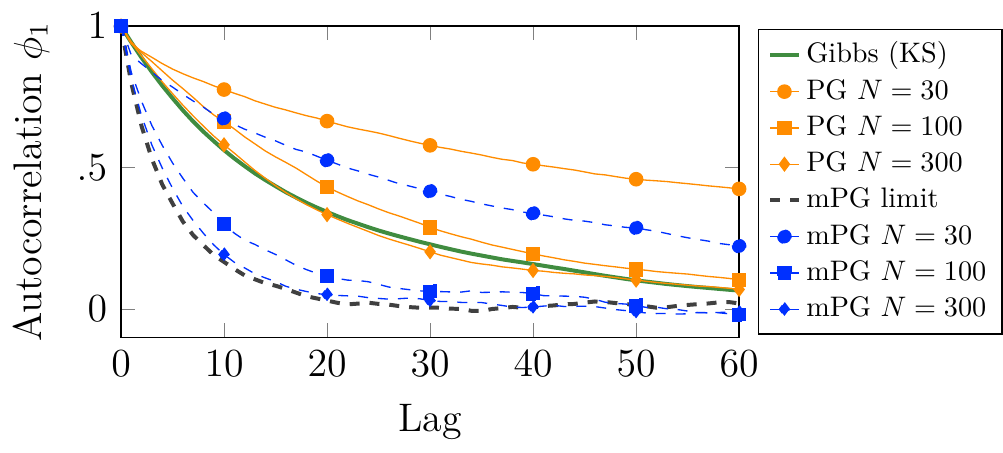}
	\caption{Autocorrelation function (ACF) for the parameter $\varT_1$. The ACF of PG converges to that of a Gibbs sampler using a Kalman smoother, whereas the single marginalized PG sampler in \eqref{eq:mPGforHSSM} converges to a different, but lower limiting ACF as $N$ increases. The limiting ACF of marginalized PG will not drop to zero due to the increased correlation from updating one dataset at a time.}
	\label{fig:acfLG3}
\end{figure}

Figure \ref{fig:acfLG3} shows the autocorrelation function\footnote{The traces were centered around the mean value for a Gibbs sampler, run for 200000 iterations, that used a Kalman smoother to produce state trajectory samples.} (ACF) for the parameter $\varT_1$ for both samplers for a varying number of particles. We also include the tractable Gibbs sampler corresponding to \eqref{eq:PGforHSSM}, realized by running a Kalman smoother for the state update; and the intractable limiting ACF of the marginalized Gibbs sampler corresponding to \eqref{eq:mPGforHSSM}, approximated using a run of single marginalized PG with $N=30000$ particles. The standard PG sampler indeed approaches the ACF of the Gibbs sampler for increasing values on $N$, but can never perform better. The single marginalized PG sampler, on the other hand, performs better than the Gibbs sampler already for $N=100$. Its limiting ACF lies well beneath the limiting ACF for the PG sampler, but it does not, in contrast to the sampler presented by \cite{Wigren2019}, yield iid samples when $N\to\infty$. This is due to the correlation introduced when updating one dataset at a time. Corresponding results for $\varT_2$, $\varO_1$ and $\varO_2$ can be found in Section \ref{app:Additional_LG} of the supplementary material. 

\begin{figure}[ht]
	\centering
	\includegraphics[width=.55\textwidth]{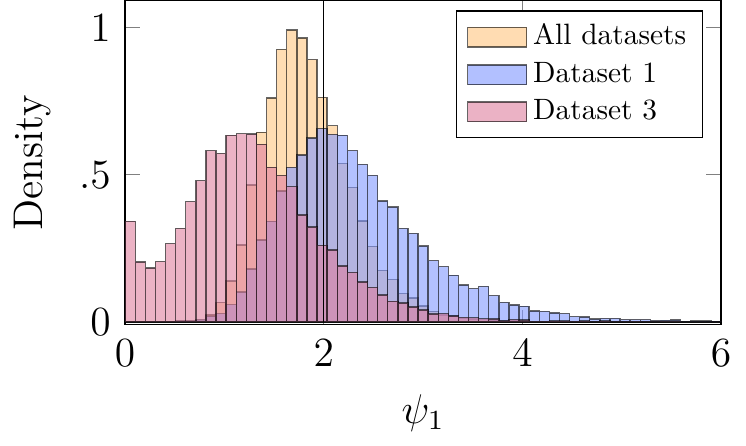}
	\caption{Histogram for the parameter $\varO_1$ obtained from running a single marginalized PG sampler, both for the multi-SSM that combines all datasets and for the case when each dataset is treated separately. The sampler that can access all datasets yields a more concentrated posterior that lies between the posteriors obtained from only using dataset 1 or 3. The solid black line indicates the true parameter value.}
	\label{fig:Histogram_3datasets_R1}
\end{figure}

Next, we illustrate the benefit of using the multi-SSM formulation rather than individual SSMs when some of the model parameters are shared between datasets. We generate data from the same multi-SSM as above and run the same single marginalized PG sampler, but using $N=100$ particles both for the multi-SSM formulation and for the case when dataset~1-3 are treated as if they were independent\footnote{When $L=1$ the single marginalized PG sampler is equivalent to the marginalized sampler by \cite{Wigren2019}.}. Figure \ref{fig:Histogram_3datasets_R1} shows the histogram for the parameter $\varO_1$ when it has been estimated using all datasets as opposed to having used only dataset~1 or dataset~3. The parameter posterior when using all datasets for estimation is narrower and lies between the parameter posteriors obtained using only dataset~1 or~3, which is what we expect to see since the multi-SSM has access to data from all three datasets. Corresponding histograms for $\varT_1$, $\varT_2$ and $\varO_2$ are available in Section \ref{app:Additional_LG} of the supplementary material. Note that the shape of the histograms is independent of the sampling algorithm used --- more or less identical histograms can be obtained from running a standard PG sampler (provided enough particles are used) or a Gibbs sampler that uses a Kalman smoother for the state updates.

\subsection{Stacked Marginalized Particle Gibbs for Multi-SSMs} \label{sec:stack}
Contrasting the approach with individual marginalized updates of all datasets is the stacking approach, where the multi-SSM is combined into one SSM by stacking the datasets together. The states are $\mathbf{X}=\left( X^1,X^2,\dots,X^L \right)$ in the stacked model, and the observations are $\mathbf{Y}=\left( Y^1,Y^2,\dots,Y^L \right)$. A marginalized PG sampler for the stacked multi-SSM iterates, after initialization at $m=1$,
\begin{equation} \label{eq:mPGstack}
\mathbf{X}(m) \sim \kappa \big(\mathbf{X}(m-1),\mathbf{X}\big),
\end{equation}
where $\kappa$ is a marginalized SMC-based Markov kernel with stationary distribution $p \big(\mathbf{X} | \mathbf{Y} \big)$. In contrast to \eqref{eq:mPGforHSSM}, this sampler updates all state trajectories simultaneously, implying that it can yield independent samples provided that enough particles are used. The drawback is that the stacked model can become high-dimensional (in time or state) and therefore requires a larger number of particles to avoid degeneracy. The stacking of the multi-SSM into a single SSM can be done in different ways. The two ways we consider here are stacking in the time dimension and stacking in the state dimension, but combinations of these approaches are also possible. 

\begin{figure}
	\centering
	\includegraphics[width=\textwidth]{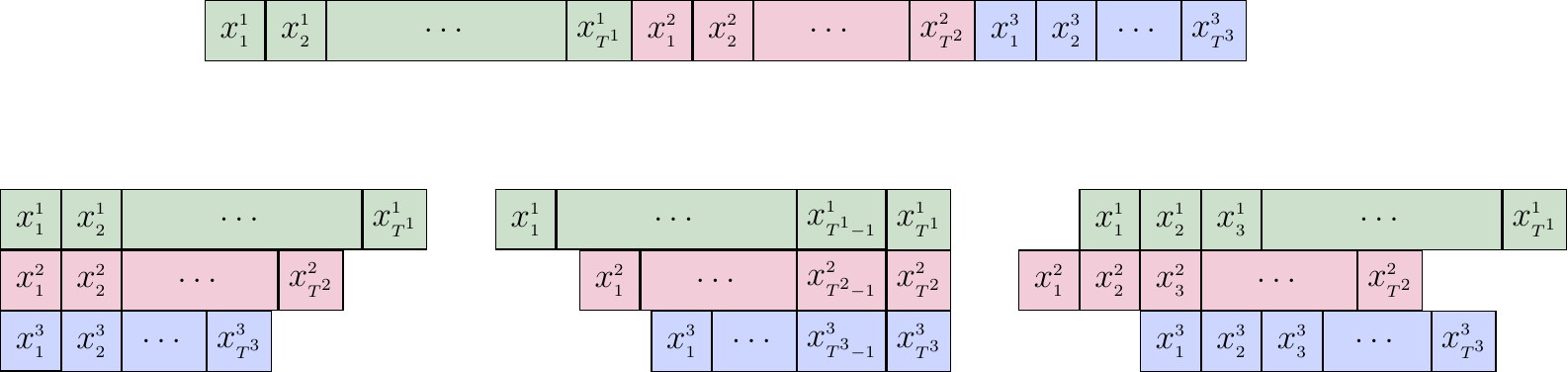}
	\caption{Different ways to stack a multi-SSM consisting of three datasets of different lengths. The stacking can be done with respect to the time dimension (top) or with respect to the state dimension (bottom). Stacking in the state dimension can, in turn, be done in different ways. The most straightforward is to either left-center (bottom left) or right-center (bottom middle) the datasets, but more sophisticated schemes where datasets overlap without being left- or right-centered are also possible (bottom right).}
	\label{fig:stacking}
\end{figure}

The simplest approach is to stack the multi-SSM with respect to time, i.e.\ form one SSM where state trajectories, and corresponding observations, are connected after each other in a sequence as illustrated in the top row of Figure \ref{fig:stacking} for the case of three datasets. A marginalized PG sampler for the stacked (in time) multi-SSM can be based on the sampler in Algorithm \ref{alg:multiSSM_mcSMC}, but the multi-SSM structure implies the following modifications to the marginalized cSMC kernel in the transition between datasets $j$ and $j+1$:
\begin{itemize}
	\item Each particle in the last timestep for dataset $j$ is the ancestor of one particle in the first timestep for dataset $j+1$.  
	\item The weights from the final timestep for dataset $j$ are initial weights for dataset $j+1$ in the sequence, i.e.\ $\bar{w}_0^{j+1}=\bar{w}_{T^j}^j$.
	\item The initial hyperparameters for dataset $j+1$ for parameters that are shared with other datasets are updated with the sum of statistics from datasets processed \textit{earlier} in the sequence, i.e.\ dataset $1$ to $j$. 
\end{itemize}
Stacking in time results in longer state trajectories that make the sampler more prone to path degeneracy, i.e.\ early timesteps have fewer effective samples than those later in the sequence. To ensure that the impoverishment for early timesteps does not affect the same dataset in every PG iteration we suggest rotating the order in which the datasets are updated between PG iterations --- either according to some chosen pattern, e.g.\ placing the first dataset last in the next update, or by picking the order of datasets randomly in each iteration.

A more intricate technique is to stack the multi-SSM in the state dimension, i.e.\ form one SSM by stacking the datasets and corresponding state trajectories ``on top of each other''. This requires syncing the datasets in time in some way, even though they may be of different lengths and stem from different processes that are not necessarily synced with respect to time\footnote{Stacking with respect to states may seem more ``natural'' when the datasets share the time scale, but it is by no means required.}. The most straightforward approaches are to either left- or right-center the datasets, illustrated in the bottom left and middle of Figure \ref{fig:stacking} for the case of three datasets. The former allows for using uninformed initial hyperparameters for all datasets, but may generate inferior samples for short datasets due to these not being updated in the last iterations. The latter leads to informed initial hyperparameters for short datasets, but allows for a greater path diversity for the later timesteps for all datasets. It is also possible to sync the datasets with respect to some other point in time, illustrated in the bottom right of Figure \ref{fig:stacking}. Two examples of when such syncing could be used are in disease modeling, where syncing with respect to the maxima of outbreaks could be beneficial, and models where a subset of the datasets may require informed initial hyperparameters to achieve good mixing of the PG sampler. 

When datasets of different lengths are stacked (in state), the state of the resulting SSM changes size across the time iterations. Fortunately, this is not a problem for the SMC algorithm as long as the proposal and weighting steps are adjusted accordingly \citep{Naesseth2019}. Irrespective of how the datasets are synced in time, marginalizing the stacked SSM introduces dependencies between the state variables that must be taken into account when forming the transition and observation densities required for the propagation and weighting steps in the marginalized cSMC kernel. To clearly illustrate the mechanics of propagation and weighting of stacked marginalized PG, we examine the special case of using a bootstrap proposal, i.e.\ the proposal distribution is the marginalized transition density and the weights are given by the marginalized observation density, and left-centered syncing in time. The extension to other proposals and time syncing is straightforward. The marginalized transition density can always be factorized according to
\begin{align} \label{eq:stackTrans}
p(&\mathbf{X}_t|\mathbf{X}_{1:t-1})=\allowbreak p(x_t^1,\dots,x_t^L|\mathbf{X}_{1:t-1})\nonumber \\ 
&= p(x_t^L|x_t^1,\dots,x_t^{L-1},\mathbf{X}_{1:t-1}) p(x_t^{L-1}|x_t^1,\dots,x_t^{L-2},\mathbf{X}_{1:t-1})
\dots p(x_t^1|\mathbf{X}_{1:t-1}),
\end{align}
which implies that the propagation step can be implemented by sampling new states from these $L$ densities in turn, starting with $x^1_t$. Note that if dataset $j$ does not share any parameters with dataset $i$, then $x_t^i$ can be removed from the conditioning for dataset $j$. The dependencies between the datasets in the marginalized transition density are manifested through the statistics, therefore these must be updated in each of the $L$ propagation steps at time $t$.
The observation density can be factorized in a similar fashion to
\begin{align} \label{eq:stackObs}
p(&\mathbf{Y}_t|\mathbf{X}_{1:t},\mathbf{Y}_{1:t-1})= p(y_t^1,\dots,y_t^L|\mathbf{X}_{1:t},\mathbf{Y}_{1:t-1}) =  p(y_t^L|y_t^1,\dots,y_t^{L-1},\mathbf{X}_{1:t},\mathbf{Y}_{1:t-1}) \nonumber \\ &\cdot p(y_t^{L-1}|y_t^1,\dots,y_t^{L-2},\mathbf{X}_{1:t},\mathbf{Y}_{1:t-1}) 
\dots p(y_t^1|\mathbf{X}_{1:t},\mathbf{Y}_{1:t-1}).
\end{align}
Thus, the weighting step can be implemented by evaluating the above densities one at a time while simultaneously updating the corresponding statistics. The marginalized cSMC algorithm for the stacked (in state) multi-SSM is similar to Algorithm \ref{alg:multiSSM_mcSMC}, but both the propagation, the weighting step and the update of statistics for transition and observation parameters are done in a loop over all datasets in each timestep.

\subsubsection{Illustration: Comparison Between Single and Stacked Marginalized Particle Gibbs} \label{sec:NeuripsCompare}
In this example we compare the performance of a stacked (in state) marginalized PG sampler with the performance of the single marginalized PG sampler from Section \ref{sec:single} and the PG sampler in \eqref{eq:PGforHSSM}. An example utilizing the marginalized PG sampler stacked with respect to time is available in Section~\ref{sec:VBD}.

\begin{figure}
	\begin{center}
		\includegraphics[width=.35\textwidth]{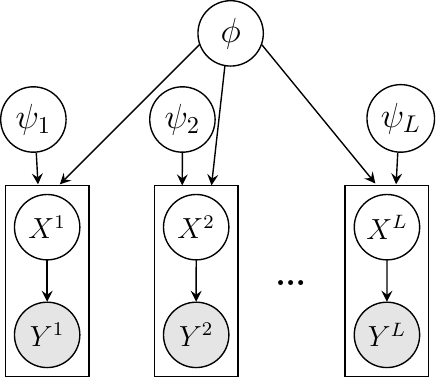}
		\caption{A linear-Gaussian multi-SSM for $L$ datasets that share process noise variance, $\varT$, but have individual observation noise variances, $\varO_\ell$.}
		\label{fig:LGdimLmodel}
	\end{center}
\end{figure}

Again, we consider a multi-SSM where the individual SSMs are linear-Gaussian with known scalar transition and observation constants, $a$ and $c$. In this example, however, all $L$ datasets have individual observation noise variances,~$\varO_\ell$, but share transition noise variance,~$\varT$. Figure \ref{fig:LGdimLmodel} depicts the structure of the multi-SSM. For simplicity, all datasets are assumed to have equal lengths $T^\ell= 30$ and are stacked with complete overlap. 

Before presenting the comparison of samplers, we formulate and discuss the stacked marginalized PG sampler in more detail. The stacked form of the multi-SSM in Figure \ref{fig:LGdimLmodel} is
\begin{align} \label{eq:LGstack1}
p(\mathbf{X}_t| \mathbf{X}_{t-1},\Theta) &= \mathcal{N}(\mathbf{X}_t| A\mathbf{X}_{t-1},\Sigma_{\text{tr}}), \\
p(\mathbf{Y}_t| \mathbf{X}_t,\Theta) &= \mathcal{N}(\mathbf{Y}_t| C\mathbf{X}_t,\Sigma_{\text{ob}}), \label{eq:LGstack2}
\end{align}
where $\Theta = (\varT, \varO_1,\dots,\varO_L)$, $A=aI_L$, $C=cI_L$, $\Sigma_{\text{tr}}=\varT I_L$ and $\Sigma_{\text{ob}}$ is a matrix with $(\varO_1,\dots,\varO_L)$ on the diagonal and zeros elsewhere. $I_L$ is the identity matrix of dimension $L$.

A marginalized PG sampler for the stacked model \eqref{eq:LGstack1}-\eqref{eq:LGstack2} can be derived by performing the same steps as in the derivation of the single marginalized PG sampler (see Section~\ref{app:single} in the supplementary material), but with the factorization of the transition and observation densities given in \eqref{eq:stackTrans} and \eqref{eq:stackObs}. 
The propagation step of the sampler iterates over all state components (1 to $L$) and propagates state component $\ell$ at time $t$ according to the ``bootstrap proposal'' $p(x_t^\ell|x_t^1,\dots,x_t^{\ell-1},\mathbf{X}_{1:t-1})$. Sampling from the proposal requires hyperparameters informed by the earlier propagation steps. The transition variance $\varT$ is shared among all individual SSMs, thus the hyperparameter updates are
\begin{align}
\chi^{\varT,\ell}_{t} &= \chi^\varT_0 + \sum_{j=1}^{L}\sum_{k=1}^{t-1}s_k^{x,j} + \sum_{j=1}^{\ell-1}s_t^{x,j} + s_t^{x,\ell} = \chi_{t}^{\varT,\ell-1} + s_t^{x,\ell}, \label{eq:HPstack1} \\
\nu^{\varT,\ell}_t &= \nu^\varT_0 + \sum_{j=1}^{L}\sum_{k=1}^{t-1}r_k^{x,j} + \sum_{j=1}^{\ell-1}r_t^{x,j} + r_t^{x,\ell} = \nu_{t}^{\varT,\ell-1} + 1, \label{eq:HPstack2}
\end{align}
where the superscript $\varT$ indicates that the hyperparameter updates are for the transition noise parameter and superscript $x$ indicates the statistic for the transition density. The observation variance $\varO_\ell$ is unique for each individual SSM and independent of the other datasets. This implies that the observation density \eqref{eq:stackObs} simplifies to $p(y_t^L|x^L_{1:t},y^L_{1:t-1})\allowbreak \cdot p(y_t^{L-1}|x^{L-1}_{1:t},y^{L-1}_{1:t-1})\allowbreak \dots \allowbreak p(y_t^1|x^1_{1:t},y^1_{1:t-1})$. Each of these factors can be evaluated separately and then multiplied together to form the weights at time $t$. The hyperparameter updates for $\varO_\ell$ at time $t$, required to evaluate the weights, are
\begin{align}
\chi^{\varO_\ell}_{t} &= \chi^{\varO_\ell}_0 + \sum_{k=1}^{t}s_k^{y^\ell} = \chi_{t-1} + s_t^{y^\ell}, \label{eq:HPstack3}\\
\nu^{\varO_\ell}_t &= \nu^{\varO_\ell}_0 + \sum_{k=1}^{t}r_k^{y^\ell} = \nu^{\varO_\ell}_{t-1} + 1, \label{eq:HPstack4}
\end{align}
where the superscript $\varO_\ell$ indicates for which observation noise parameter the updates are valid and superscript $y^\ell$ indicates the statistic for the observation density for dataset $\ell$.
A comparison of the expressions for the hyperparameter updates \eqref{eq:HPstack1}-\eqref{eq:HPstack4} in stacked marginalized PG with those for single marginalized PG \eqref{eq:hyperParam1}-\eqref{eq:hyperParam2} shows that the former only uses information from the other datasets up to time $t-1$ when updating the state at time $t$, whereas the latter takes all timesteps from the other datasets into account. 

To compare the performance of PG, single marginalized PG and stacked marginalized PG, we consider the \textit{improvement factor} relative to PG, i.e.\ the quotient between the integrated ACF of the PG sampler and the marginalized method.
We run all algorithms for $M=15000$ iterations, with 1500 samples discarded as burn-in, for multi-SSMs that consists of 1 up to 12 datasets. 
For all number of datasets, we use $N=100$ particles for single marginalized PG and $N=200$ particles for PG and stacked marginalized PG to ensure comparable running times between methods.
Additional details about the experiment are available in Section~\ref{app:impDetailLG} of the supplementary material.

Figure \ref{fig:impFact} shows the mean improvement factor for $\varT$ over five runs.
A value above one indicates that the marginalized method performs better than PG. 
Stacked marginalized PG is the most efficient method for up to four datasets. For increasing $L$ it becomes more and more degenerate due to an insufficient number of particles to representing the high-dimensional state trajectory, which leads to weight degeneracy.
The single marginalized PG sampler updates one state trajectory at a time and is therefore not affected by the curse of dimensionality to the same extent. It always performs better than PG, but the difference decreases with an increasing number of datasets. 
It should be noted that for $L=1$, the two marginalized methods are both equivalent to the marginalized sampler described by \cite{Wigren2019}. The apparent difference between them in Figure \ref{fig:impFact} is due to randomness and the different number of particles used. Furthermore, it should be emphasized that in this example the two marginalized methods performed equally good for four datasets, but this should not be interpreted as a general rule. The intersection will occur at a different number of datasets for different multi-SSMs and also depends on the number of particles used in the cSMC kernels in relation to the length of the datasets --- even for the same multi-SSM it can vary between runs.

\begin{figure}[ht]
	\centering
	\includegraphics[width=.55\textwidth]{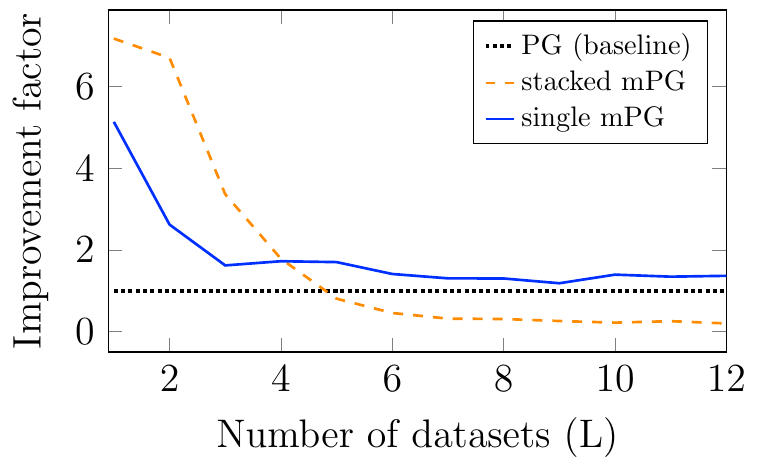}
	\caption{Improvement factor with respect to the PG sampler for the transition noise variance, $\varT$, over five runs. The improvement factor is the quotient between the integrated autocorrelation function of the alternative method and the PG sampler.
		The single marginalized PG sampler performs better than the PG sampler for all number of datasets. The stacked marginalized PG sampler excel for a small number of datasets, but degenerate as the number of datasets increases.
		To make running times comparable between methods, PG and stacked marginalized PG were run with twice as many particles in the cSMC part. }
	\label{fig:impFact}
\end{figure}

An important property of the stacked marginalized PG sampler is its ability to yield (close to) iid samples. In Section \ref{app:Additional_LGstack} of the supplementary material we provide a small example that illustrates this property for the case of four datasets with a shared transition noise parameter.

\subsection{State Trajectory Diversity in the Marginalized Samplers}
The correlation properties of the marginalized PG samplers are tightly linked to how the dependencies arising from marginalization between datasets with shared parameters are managed in the samplers. 
In the single marginalized PG sampler, the influence from other datasets on a state trajectory update stems from a \textit{single} reference trajectory from each dataset, which affects the initial hyperparameters for \textit{all} particles in the SMC-based kernel. In particular for longer datasets, the informed initial hyperparameters (identical for all particles) can cause the marginalized transition and observation densities to become very narrow, which in turn impedes the generation of state trajectories that are significantly different from the reference trajectory, leading to highly correlated state trajectories and poor mixing of the sampler. 
For the two stacked samplers, the impact from other datasets in each update is more intricate. In the marginalized PG sampler stacked in time, the update of the first state trajectory in the sequence is not affected by other datasets, but succeeding state trajectories are increasingly affected by informative initial hyperparameters the further back in the sequence they are located. An important distinction from the restrictive informative hyperparameters in the single marginalized PG sampler is that there are $N$ \textit{different} informative hyperparameters when the update of the next dataset starts, which contributes to more diverse state trajectories and, in turn better mixing, provided that path degeneracy is avoided. 
In the marginalized PG sampler stacked with respect to the state dimension, the influence from other datasets depends on how they are synced in time. For simplicity, consider the case of left-centering (the mechanics are the same for other syncings). In consecutive timesteps, statistics are accumulated for one dataset at a time, implying that the informative hyperparameters only consist of information from other datasets \textit{up until} that timestep and there are $N$ \textit{different} informative hyperparameters (one for each particle). This yields more diverse state trajectories, which eases selecting a trajectory that differs from the reference trajectory.
It should be noted that by shifting the time-syncing of the stacked model we can let some datasets be influenced by statistics from the other datasets to a varying degree. The extreme case is when there is no overlap between datasets, which is equivalent to the stacked in time marginalized PG sampler.

\section{Simulation Study: Inference in Mosquito-borne \\ Disease Models} \label{sec:VBD}
In this section we consider the performance of a marginalized PG sampler on a non-linear, non-Gaussian multi-SSM describing the spread of two mosquito-borne diseases during three different outbreaks. Besides showing that the marginalized framework is useful for more complicated models with real-world data, the simulation study will also illustrate how to design a multi-SSM given attributes of the underlying process and how to choose a suitable structure for the marginalized PG sampler.

\textit{Dengue fever} is a mosquito-borne, viral disease endemic to tropical areas. Although a majority of the estimated 100 to 400 million cases each year are mild or even asymptomatic, around 24 million of them cause hospitalizations and 10000 cause death \citep{Bhatt2013,WHO2022}.
\textit{Zika} is a milder viral disease where the vast majority of the infected individuals are asymptomatic. It is spread by the same species of mosquito as dengue fever, the Aedes aegypti, and is therefore present in the same areas. No deaths have been reported, but recent studies have shown that Zika can cause children to be born with microcephaly if the mother is infected during pregnancy \citep{WHO2018}.  

We use three datasets from outbreaks of dengue and Zika virus in Micronesia originally presented by \cite{Funk2016}\footnote{The datasets are available as supplementary material of \cite{Funk2016}.}. Each dataset consists of daily or weekly observations of the number of newly infected humans that visited a health center. The first dataset contains 16 weekly observations of the number of new human infections during an outbreak of Zika virus on Yap in 2007\footnote{This was, in fact, the first known major outbreak of Zika in a human population.}. The second dataset is from an outbreak of dengue fever on Yap in 2011 and contains 183 daily observations of the number of newly infected humans. The third dataset contains 42 daily observations of newly infected humans from an outbreak of dengue fever on the smaller island Fais in 2011. The study by \cite{Funk2016} used a model structure with shared parameters to learn more about the less documented transmission process of the Zika virus from the dengue outbreaks. Learning about new diseases through outbreaks of better known, related diseases is another motivation for studying inference methods for multi-SSMs.

\subsection{Designing Multi-SSMs for Mosquito-borne Disease Outbreaks} 
\begin{figure}
	\centering
	\includegraphics[width=.95\textwidth]{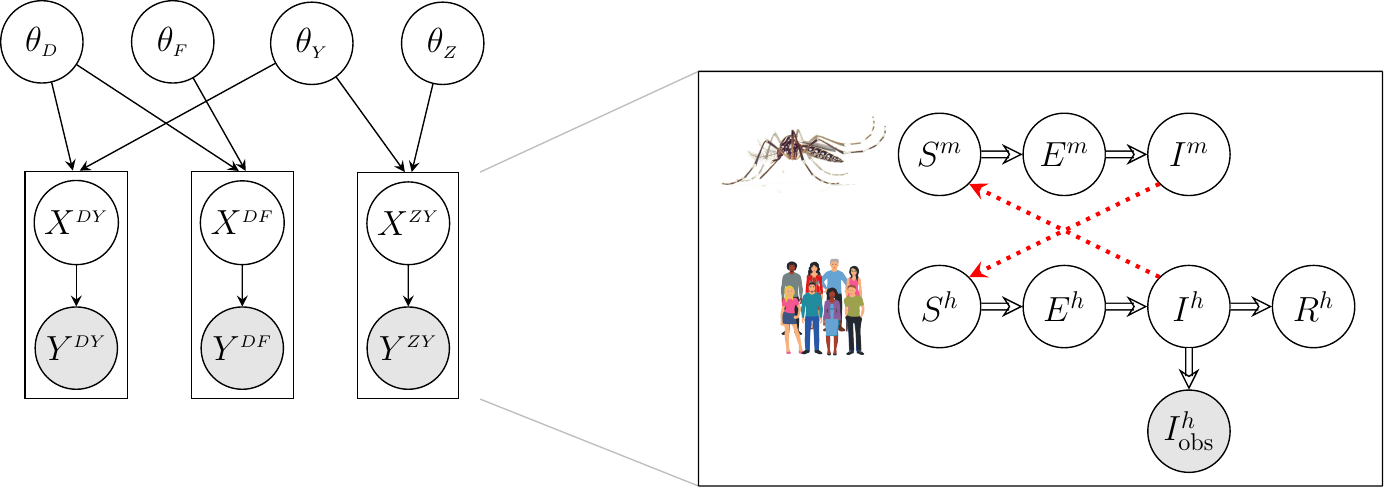}
	\caption{The multi-SSM describing the dengue and Zika outbreaks (left) and the compartmental model for a single dataset (right). The outbreak of dengue on Yap (leftmost sub-SSM) shares parameters with all outbreaks, whereas the Zika outbreak and the outbreak on Fais have no common parameters. The compartmental model illustrates the stochastic flow of individuals between compartments (double arrows) and how the mosquito and human models are coupled through cross-infection (red dashed arrows). The number of individuals in each of the compartments $S$, $E$, $I$ and $R$ constitute a part of the state trajectories $X$ in the multi-SSM, and the number of newly infected humans, $I_\text{obs}^h$ are the observations $Y$.}
	\label{fig:VBDmodel}
\end{figure}

To construct the multi-SSM, we first build the large-scale structure, i.e.\ relate parameters (or groups of parameters) to datasets, and then consider the individual SSMs in more detail. The two dengue datasets contain data from outbreaks of the same disease and can therefore be assumed to share \textit{disease-specific} parameters, like the incubation or infection time, that are the same independent of where in the world the outbreak occurs.
Dengue and Zika virus are both transmitted by the Aedes aegypti mosquito. It is therefore reasonable to assume that parameters related to the vector, such as the mosquito density or the bite frequency, are \textit{location-specific} and shared between the two outbreaks on Yap.
Based on the above assumptions about the parameters, the dengue outbreak on Yap shares parameters with both the dengue outbreak on Fais and the Zika outbreak on Yap, while the Fais outbreak have no parameters in common with the Zika outbreak. The resulting multi-SSM is depicted in Figure \ref{fig:VBDmodel} (left). We use indices $D$ and $Z$ to denote the disease (dengue or Zika), and indices $Y$ and $F$ to denote the location (Yap or Fais). Thus, $X^{\sss{DF}}$ is the state trajectory for the dengue outbreak on Fais, $Y^{\sss{ZY}}$ are the observations from the Zika outbreak on Yap etc.  

Each individual SSM in Figure \ref{fig:VBDmodel} consists of two coupled compartmental models \citep{Kermack1927,Smith2012}: a susceptible-exposed-infectious-recovered (SEIR) model that describes the transitions between compartments for the human population and a susceptible-exposed-infectious (SEI) model that describes transitions between compartments for the mosquito population. The coupling between the two compartmental models stems from the fact that infectious mosquitoes can infect susceptible humans and vice versa. The compartmental structure is illustrated in Figure \ref{fig:VBDmodel} (right). 

We employ a discrete, stochastic formulation of the compartmental model based on that described by \cite{Murray2018}\footnote{The model by \cite{Murray2018} is in turn based on the continuous, deterministic model used by \cite{Funk2016}.}, where transitions between compartments are binomially distributed with beta priors on the unknown parameters to allow for marginalization. 
However, the model by \cite{Murray2018} does not incorporate any location dependent parameter that can be marginalized out. Therefore, we have modified the susceptible-to-exposed transition to include a location dependent bite frequency, $b$. It is assumed to form a Poisson-gamma conjugacy relation with the number of interactions between susceptible and infectious individuals to allow for marginalization. 
Given the number of interactions between susceptible and infectious individuals, the number of newly exposed individuals can then be drawn from the \textit{occupancy distribution} \citep{ONeill2021}. 
A detailed motivation for this modification and a complete model description, including other minor modifications to the model by \cite{Murray2018}, is available in Section~\ref{app:VBDtrans}-\ref{app:VBDparam} of the supplementary material. 
To account for the uncertain proportion of cases reported to health centers, the observations are assumed to follow a binomial distribution with an unknown reporting probability parameter. 
A beta prior is assumed for the reporting probability $\rho$.

\subsection{Designing the Marginalized Particle Gibbs Sampler} 
An important aspect when designing the marginalized PG sampler is the dimension of the state space. Each individual SSM is fairly high-dimensional with ten states and seven parameters. However, the observations are one-dimensional for each dataset and there are only three datasets, thus, at least in theory, both marginalized samplers presented in Section \ref{sec:multiSSM_mPG} are viable options. 
Based on the above, the single marginalized PG sampler is an attractive first choice, due to its simplicity implementation-wise and its ability to keep the dimension of the SMC kernel low. 
However, the coupled compartmental model has strong dependencies---both between parameters and between states and parameters---which yield too informative hyperparmeters when conditioning on the other datasets, resulting in poor mixing.
In an attempt to loosen these dependencies we use a marginalized PG sampler that updates the datasets that share location together, and the datasets that share disease together to avoid (or at least reduce) the influence from the strongly correlated dataset on the hyperparameters.
This sampler can be viewed as a hybrid between the single and stacked marginalized samplers, in the spirit of a partially collapsed Gibbs sampler \citep{vanDyk2008}. 
Since there are only three datasets of one-dimensional observations, stacking with respect to the state dimension is possible, but it is not clear how to best sync the datasets with respect to time when the sampling times differ (daily and weekly data). To avoid this issue we stack the datasets with respect to time instead, which is also simpler implementation-wise.
To ensure that path degeneracy does not affect the same state trajectory in every update we switch the order that the state trajectories are updated in every second iteration, a design choice worked well for this particular model. To summarize, the updates during two iterations of the marginalized PG sampler, after initialization, are
\begin{align} 
&\text{if } m\text{ even} \nonumber \\
&\hspace{7mm} X^{\sss{DF}}(m),X^{\sss{DY}}(m) \sim \kappa^{\sss{D_e}}\Big( \{X^{\sss{DF}}(m-1), X^{\sss{DY}}(m-1)\}, \{X^{\sss{DF}}, X^{\sss{DY}}\}\Big), \label{eq:kappaDenE}\\
&\hspace{7mm} X^{\sss{DY}}(m),X^{\sss{ZY}}(m) \sim \kappa^{\sss{Y_e}}\Big(\{X^{\sss{DY}}(m),X^{\sss{ZY}}(m-1)\},\{X^{\sss{DY}},X^{\sss{ZY}}\}\Big) \label{eq:kappaYapE}\\
&\text{if } m\text{ odd} \nonumber \\
&\hspace{7mm} X^{\sss{DY}}(m),X^{\sss{DF}}(m) \sim \kappa^{\sss{D_o}}\Big( \{X^{\sss{DY}}(m-1), X^{\sss{DF}}(m-1)\}, \{X^{\sss{DY}}, X^{\sss{DF}}\}\Big), \label{eq:kappaDenO}\\
&\hspace{7mm} X^{\sss{ZY}}(m),X^{\sss{DY}}(m) \sim \kappa^{\sss{Y_o}}\Big(\{X^{\sss{ZY}}(m-1),X^{\sss{DY}}(m)\},\{X^{\sss{ZY}},X^{\sss{DY}}\}\Big) \label{eq:kappaYapO}
\end{align}
where $\kappa^{\sss{D_e}}$ and $\kappa^{\sss{D_o}}$ are marginalized SMC-based Markov kernels with stationary distribution $p(X^{\sss{DY}},X^{\sss{DF}}|Y^{\sss{DY}},Y^{\sss{DF}},\mathcal{D}^{\sss{ZY}})$ that updates the dengue state trajectories, and $\kappa^{\sss{Y_e}}$ and $\kappa^{\sss{Y_o}}$ are marginalized SMC-based Markov kernels with $p(X^{\sss{DY}},X^{\sss{ZY}}|Y^{\sss{DY}},Y^{\sss{ZY}},\mathcal{D}^{\sss{DF}})$ as stationary distribution that updates the Yap state trajectories. 
Note that once we have updated the dengue on Yap state trajectory in the dengue updates, \eqref{eq:kappaDenE} and \eqref{eq:kappaDenO}, we need to use the new trajectory in the updates of the Yap state trajectories, \eqref{eq:kappaYapE} and \eqref{eq:kappaYapO}.
The difference between the even and odd updates is the order of the temporal stacking of trajectories.

\subsection{Simulation Results}
We compare the performance of the marginalized PG sampler in \eqref{eq:kappaDenE}-\eqref{eq:kappaYapO} with that of the standard PG sampler in \eqref{eq:PGforHSSM} for the multi-SSM in Figure \ref{fig:VBDmodel}. We also compare the marginalized PG sampler with separate marginalized runs for each dataset using Algorithm \ref{alg:multiSSM_mcSMC}. All simulations were run for 5000 MCMC iterations, where the first 500 samples were discarded as burn-in. The marginalized PG sampler used 7000 particles for both the disease update and the location update. The standard PG sampler and the individual marginalized runs used 10000 particles for the dengue on Yap dataset, 4000 particles for the dengue on Fais dataset and 5000 particles for the Zika on Yap dataset. Initial experiments showed that even for a much larger number of particles than those used in the final experiments, the standard PG sampler exhibited poor mixing implying that the bottleneck is the strong dependencies in the multi-SSM. Additional implementation details are available in Section \ref{app:VBDimplement} of the supplementary material.

\begin{figure}[ht]
	\centering
	\includegraphics[width=.85\textwidth]{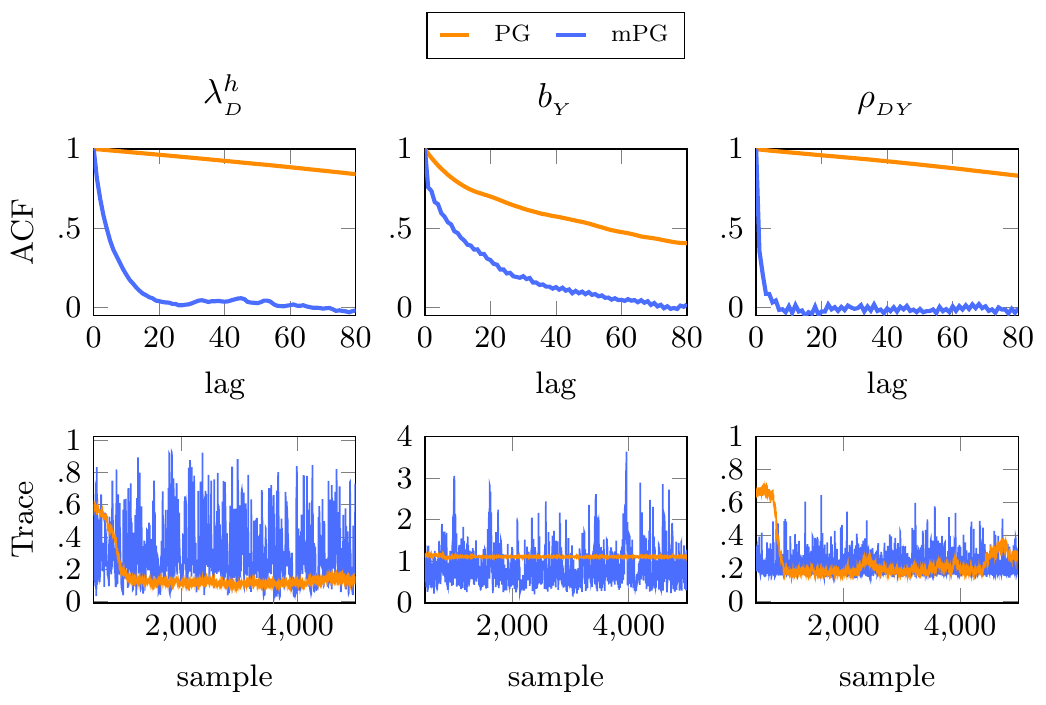}
	\caption{ACF and trace plot for a disease dependent parameter ($\lambda^h_{_D}$), a location dependent parameter ($b_{_Y}$) and an outbreak dependent parameter ($\rho_{_{DY}}$). While the standard PG sampler experiences poor mixing, the marginalized PG sampler mixes well for all three parameters. Corresponding histograms for the marginalized PG sampler are shown in Figure \ref{fig:Histogram_VBD_1vs3} (the PG sampler is excluded due to poor mixing).}
	\label{fig:PGvsmPG_VBD}
\end{figure}

We start by evaluating the benefit of using the marginalized PG sampler \eqref{eq:kappaDenE}-\eqref{eq:kappaYapO} in relation to the standard PG sampler in \eqref{eq:PGforHSSM}.
Figure \ref{fig:PGvsmPG_VBD} shows the ACF and trace plot for three of the model parameters: the disease dependent parameter $\lambda_{_D}^h$ (transmission probability for humans), the location dependent parameter $b_{_Y}$ (bite frequency) and the dataset dependent parameter $\rho_{_{DY}}$ (reporting probability). It is clear that the PG sampler mixes poorly and produces highly correlated samples for all three parameters. The marginalized PG sampler mixes well and yields samples with a much weaker correlation. Corresponding results for the remaining parameters are available in Section~\ref{app:Additional_VBD} of the supplementary material.

\begin{figure}[ht]
	\centering
	\includegraphics[width=.95\textwidth]{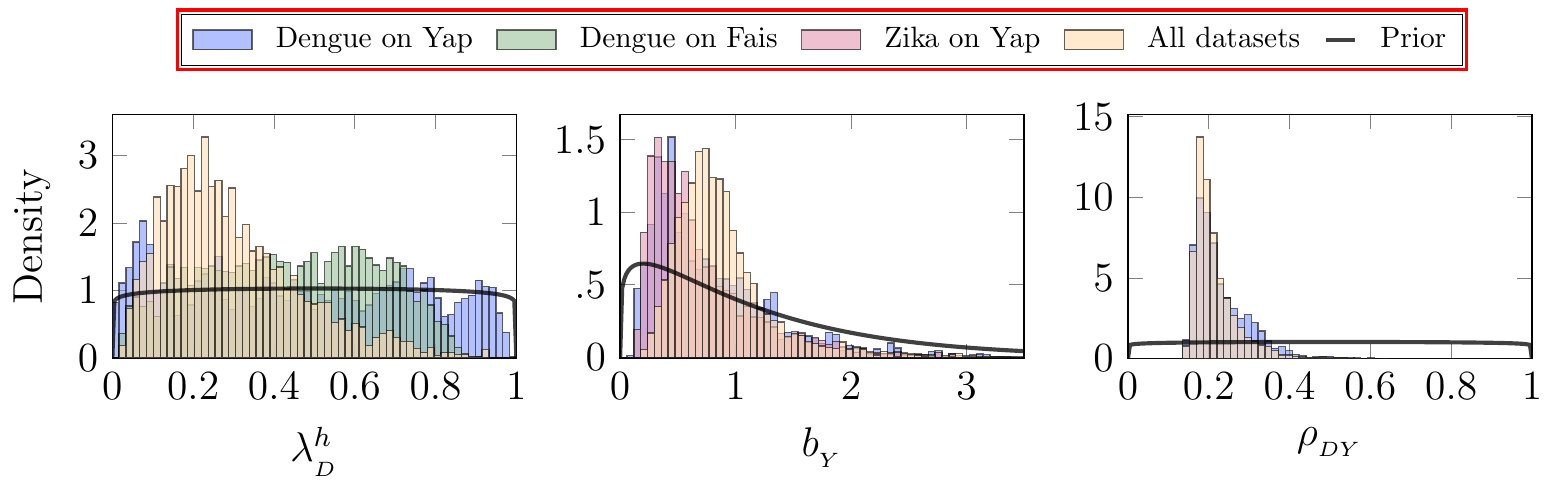}
	\caption{Histograms for the parameters $\lambda^h_{_D}$, $b_{_Y}$ and $\rho_{_{DY}}$ obtained from running the marginalized PG sampler, both for the multi-SSM that uses all datasets for parameter estimation and for the case when the datasets are treated separately.}
	\label{fig:Histogram_VBD_1vs3}
\end{figure}

We also examine the potential gain from performing inference in the multi-SSM setting rather than treating each dataset individually. 
Figure \ref{fig:Histogram_VBD_1vs3} shows the parameter posteriors for the three parameters, $\lambda_{_D}^h$, $b_{_Y}$ and $\rho_{_{DY}}$, both when using a single dataset and when using all datasets for inference. For $\lambda_{_D}^h$ and $b_{_Y}$ the multi-SSM formulation results in posterior samples further from the prior than when just using a single dataset for inference. For $\rho_{_{DY}}$, the two histograms are very alike, which is reasonable since $\rho$ is a outbreak-specific parameter. The posteriors for the multi-SSM formulation are narrower than the posteriors from a single dataset for all three parameters, likely due to the availability of more information in the multi-SSM formulation. Histograms for the remaining model parameters are available in Section \ref{app:Additional_VBD}.

\section{Discussion} \label{sec:Discussion}
PG samplers are attractive and widely used methods for Bayesian inference, in part due to their ease of implementation. They often perform well \citep{Calafat2018,Linderman2014,Meent2015,Valera2015}, but struggle for models with strong dependencies between states and parameters. For multi-SSMs --- a model structure where multiple SSMs are connected by shared parameters --- already strong dependencies may be emphasized further through the parameter sharing.
The performance of PG samplers for single SSMs can, in many cases, be significantly improved by marginalizing out the parameters from the state update \citep{Wigren2019}. Building on those results to improve the performance of PG samplers for multi-SSMs, we have described two contrasting marginalization strategies applicable to multi-SSMs --- single and stacked marginalized PG. We have also shown that they can be combined to form a marginalized PG sampler that performs well for a more complicated model describing disease outbreaks, where both base marginalization strategies used on their own struggle. 

In fact, the single and stacked marginalization strategies can be combined in a multitude of ways to form new marginalized PG samplers with different properties. The most efficient sampler design is likely model dependent and developing general guidelines for how to construct the sampler given a model structure is an interesting topic for future work. However, based on the experiments in this article some initial pointers regarding the design process can be given. Because it is simple to implement and does not involve any design choices, a good starting-point is the single marginalized PG sampler, in particular if the multi-SSM contains many or long datasets. If the sampler mixes poorly and the multi-SSM does not contain too many or too long datasets, one of the stacked samplers should be considered next. Additional design choices then include how to sync the datasets in time if the stacking is with respect to the state dimension, and how to interchange the ordering of the state trajectories between iterations if the stacking is with respect to the time dimension. 
Finally, if the correlation between two or more of the datasets is strong and the stacked methods are too computationally expensive, it can be a good idea to do a blocked (stacked) update of the strongly correlated state trajectories while updating the remaining ones using the single marginalized PG sampler. Furthermore, it may be possible to incorporate marginalized versions of other SMC-based Markov kernels designed to handle high-dimensional models, such as the one by \cite{Finke2021}.

How to best sync the datasets in time when applying the stacked (in state) marginalized PG sampler is an open question. The assumption in the multi-SSM is that the datasets have been generated from different processes, independent of each other. Consequently, it should be possible to sync the datasets ``as we like''. Based on the type of inference method used, there could be a benefit from syncing in certain ways to, e.g.\, avoid path degeneracy in the SMC-based kernels for shorter datasets (right-sync) or to ensure that the influence from other datasets is weak (left-sync). 
Likewise, more work is needed to evaluate how to order the datasets when stacking them with respect to time.

The marginalized framework requires conjugacy relations between parameter priors and observation/transition densities, which can seem restrictive in practice.
However, the marginalized PG samplers presented in this article are all valid MCMC kernels and can therefore be combined with other MCMC kernels to form an algorithm that marginalizes out all conjugate parameters and updates the remaining parameters (with no conjugacy relation) using, e.g.\, a Metropolis-Hastings sampler. 

Advances in probabilistic programming during the past years have contributed to making Bayesian inference methods accessible to a larger audience by providing a framework where the user specifies the model and can choose from a selection of already implemented inference methods \citep{Meent2018, Murray2018a}. Some PPLs can even tune the inference method or provide analytic improvements, such as marginalization, \textit{automatically} \citep{Hoffman2014,Hoffman2018,Obermeyer2019}. 
Automated marginalized PG for a single SSM is already available in a PPL \citep{Wigren2019}; incorporating the marginalized PG samplers for multi-SSMs into a PPL is an interesting path of future work. 
The modular formulation of the single marginalized PG sampler makes it particularly suitable --- it suffices to implement Algorithm~\ref{alg:multiSSM_mcSMC} (inner module) and provide support for the multi-SSM structure and the handling of hyperparameter updates (outer module).
In the ideal case the PPL should be able to choose a suitable marginalized PG sampler (single, stacked or a combination) for a given multi-SSM, but this would require guidelines for when to use each sampler that are not in place yet. 
However, merely providing users with pre-implemented versions of single and stacked marginalized PG and the option to put together a combined marginalized sampler would be a great asset.

\bigskip

\begin{center}
	{\large\bf Acknowledgements}
\end{center}
This work was supported by the Swedish Research Council under the Grants \textit{NewLEADS - New Directions in Learning Dynamical Systems} (contract number: 621-2016-06079) and \textit{Handling Uncertainty in Machine Learning Systems} (contract number: 2020-04122); 
the Knut and Alice Wallenberg Foundation through the Wallenberg AI, Autonomous Systems and Software Program (WASP);
by the Excellence Center at Linköping--Lund in Information Technology (ELLIIT);
and by \textit{Kjell och Märta Beijer Foundation}.

The authors also thank the Swedish National Infrastructure for Computing (SNIC) at UPPMAX, partially funded by the Swedish Research Council through grant agreement no. 2018-05973, for providing computational resources for a part of the simulation study through the project SNIC 2022/22-302.

\newpage

\bibliography{refs}

\newpage

\newpage

\begin{center}
	{\huge\bf Supplementary Material}
\end{center}

\bigskip

\begin{center}
	{\Large\bf Marginalized particle Gibbs for multiple state-space models coupled through shared parameters}
\end{center}

\bigskip

\begin{minipage}{.45\textwidth}
	\begin{center}
		{\footnotesize{Anna Wigren}}\\
		{\scriptsize{Department of Information Technology,\\ Uppsala University, Sweden \\ anna.wigren@it.uu.se \\}}
	\end{center}
\end{minipage}%
\begin{minipage}{.5\textwidth}
	\begin{center}
		{\footnotesize{Fredrik Lindsten}} \\
		{\scriptsize{Department of Computer and Information Science, \\ Linköping University, Sweden \\ fredrik.lindsten@liu.se \\}}
	\end{center}
\end{minipage}

\vspace{1.5cm}
The supplementary material contains a derivation of the single marginalized particle Gibbs sampler for multi-SSMs (Section \ref{app:single}), details about the linear-Gaussian experiments (Section \ref{app:impDetailLG}), a detailed description of the model used in the simulation study and its implementation (Section \ref{app:VBD}), and additional experimental results (Section \ref{app:results}).

\appendix

\section{Derivation of the Single Marginalized PG Kernel} \label{app:single}
To derive the single marginalized SMC-based kernel $\kappa^\ell$ in \eqref{eq:mPGforHSSM}, we let its unnormalized marginalized target density be $\gamma_t(x_{0:t}^\ell) = p\big(x_{0:t}^{\ell},y_{1:t}^{\ell} | \mathcal{D}^{\neg \ell} \big)$. Provided that an appropriate proposal distribution $q_t$ is selected, we only need to derive an expression for the marginalized weights to fully specify the kernel $\kappa^\ell$. The weights \eqref{eq:SMCweight} from the general SMC formulation are, for the marginalized target, given by 
\begin{equation}
\omega_{t}(x_{0:t}^\ell) = \frac{\gamma_{t}(x_{0:t}^\ell)}{\gamma_{t-1}(x_{0:t-1}^\ell)q_{t}(x_{t}^\ell | x_{0:t-1}^\ell)}=\frac{p(x_t^\ell,y_t^\ell | x_{0:t-1}^\ell,y_{1:t-1}^\ell,\mathcal{D}^{\neg \ell})}{q_{t}(x_{t}^\ell | x_{0:t-1}^\ell)},
\end{equation}
where we have utilized
\begin{equation}
\gamma_t(x_{0:t}^\ell)=p(x_{0:t}^\ell,y_{1:t}^\ell | \mathcal{D}^{\neg \ell})=p(x_0^\ell)\prod\limits_{k=1}^{t}p(x_k^\ell,y_k^\ell | x_{0:k-1}^\ell,y_{1:k-1}^\ell,\mathcal{D}^{\neg \ell}).
\end{equation} 
The numerator in the weight expression is 
\begin{align} \label{eq:SmargInt}
p(x_t^\ell,y_t^\ell |& x_{0:t-1}^\ell,y_{1:t-1}^\ell,\mathcal{D}^{\neg \ell}) = \int p(x_t^\ell,y_t^\ell, \Theta |  x_{0:t-1}^\ell,y_{1:t-1}^\ell,\mathcal{D}^{\neg \ell})\mathrm{d}\Theta \nonumber \\ 
&=\int p(x_t^\ell,y_t^\ell | x_{t-1}^\ell,\theta^\ell)p(\Theta | x_{0:t-1}^\ell,y_{1:t-1}^\ell,\mathcal{D}^{\neg \ell})\mathrm{d}\Theta,
\end{align}
where $p(\Theta | x_{0:t-1}^\ell,y_{1:t-1}^\ell,\mathcal{D}^{\neg \ell})$ is the posterior distribution of the parameters. The second equality is due to the fact that given the parameters $\theta^\ell$ of model $\ell$ and the previous state $x_{t-1}^\ell$, the likelihood is independent of all previous states and observations and all other parameters in $\Theta$. Note also that we assume that all parameters are independent of one another, thus when (later on) integrating over the complete parameter vector $\Theta$, all factors in the parameter posterior corresponding to parameters not included in $\theta^\ell$ will integrate to one. However, for notational reasons we keep the complete parameter posterior for now. 

The complete parameter posterior is
\begin{align} \label{eq:CompletePost}
p(&\Theta|x_{0:t-1}^\ell,y_{1:t-1}^\ell,\mathcal{D}^{\neg\ell}) \propto p(x_{0:t-1}^\ell,y_{1:t-1}^\ell|\mathcal{D}^{\neg\ell},\Theta)p(\mathcal{D}^{\neg\ell}|\Theta)p(\Theta) \nonumber \\
&= p(x_{0:t-1}^\ell,y_{1:t-1}^\ell|\Theta)\prod\limits_{j\neq\ell}p(x_{0:t-1}^j,y_{1:t-1}^j|\Theta)\prod_{j=1}^L p(\theta^\ell) \nonumber \\
&= p(x_0^\ell)\prod_{k=1}^{t-1}p(x_k^\ell,y_k^\ell|x_{k-1}^\ell\theta^\ell)\prod_{j\neq\ell}p(x_0^j)\prod_{k=1}^{T^j}p(x_k^j,y_k^j|x_{k-1}^j,\theta^j)\prod_{j=1}^L p(\theta^\ell)
\end{align}
where the first equality is due to all datasets being independent given the parameters and all parameters being independent of one another. The complete likelihoods in the second equality is due to independence with respect to earlier states and observations given the state at $k-1$ and the parameters of model $j$ or $\ell$.

To evaluate \eqref{eq:CompletePost} exactly we now consider likelihoods and parameter priors in the exponential family. Inserting \eqref{eq:EfamLhood}-\eqref{eq:EfamPrior} in \eqref{eq:CompletePost} yields
\begin{align}
p(&\Theta|x_{0:t-1}^\ell,y_{1:t-1}^\ell,\mathcal{D}^{\neg\ell}) \propto 
p(x_0^\ell)\prod_{k=1}^{t-1} h_k^\ell\exp\left({\theta^{\ell}}^\Tr s_k^\ell - {A^\ell}^\Tr(\theta^\ell)r_k^\ell\right)
\prod_{j\neq\ell} p(x_0^j)\nonumber \\
&\cdot \prod_{k=1}^{T^j} h_k^j  \exp\left({\theta^{j}}^\Tr s_k^j - {A^\ell}^\Tr(\theta^j)r_k^j\right) 
\prod_{j=1}^L g^\ell(\chi_0^j,\nu^j)\exp\left( {\theta^j}^\Tr \chi_0^j-{A^\ell}^\Tr(\theta^j)\nu_0^j \right).
\end{align}
Now, to compute the weights, i.e.\ evaluate the integral \eqref{eq:SmargInt}, only the parameter posterior for the set of parameters $\theta^\ell$ is required, not the complete parameter posterior. However, $\theta^\ell$ contains a subset of the parameters in $\Theta$ and some of these parameters might be shared with some of the other model's set of parameters. Therefore, we now use the $\map{j}{\ell}$ operator, which only copies statistics corresponding to parameters from dataset $j$ that are shared with dataset $\ell$, to incorporate the (potential) influence on $\theta^\ell$ from statistics from other datasets. The parameter posterior for $\theta^\ell$ is
\begin{align} \label{eq:Th_ellPost}
p(&\theta^\ell|x_{0:t-1}^\ell,y_{1:t-1}^\ell,\mathcal{D}^{\neg\ell}) \propto 
p(x_0^\ell)\left(\prod_{k=1}^{t-1} h_k^\ell \right) \exp\left({\theta^{\ell}}^\Tr \sum_{k=1}^{t-1}s_k^\ell - {A^\ell}^\Tr(\theta^\ell)\sum_{k=1}^{t-1}r_k^\ell\right) \nonumber \\
& \cdot \left(\prod_{j\neq\ell} p(x_0^j)\prod_{k=1}^{T^j} h_k^j\right) \exp\Bigg({\theta^{j}}^\Tr \sum_{j\neq\ell}\sum_{k=1}^{T^j} \map{j}{\ell}(s_k^j) - {A^\ell}^\Tr(\theta^j)\sum_{j\neq\ell}\sum_{k=1}^{T^j} \map{j}{\ell}(r_k^j)\Bigg) \nonumber \\
&\cdot g^\ell(\chi_0^\ell,\nu^\ell)\exp\left( \theta^\ell\chi_0^\ell-{A^\ell}^\Tr(\theta^\ell)\nu_0^\ell \right) = p(x_0^\ell)g^\ell(\chi_0^\ell,\nu^\ell) \left(\prod_{k=1}^{t-1} h_k^\ell \right)\left(\prod_{j\neq\ell}p(x_0^j) \prod_{k=1}^{T^j} h_k^j\right)  \nonumber \\
& \exp\Bigg( {\theta^\ell}^\Tr \bigg( \chi_0^\ell + \sum_{j\neq\ell}\map{j}{\ell}(S^j) + \sum_{k=1}^{t-1}s_k^\ell \bigg) - {A^\ell}^\Tr(\theta^\ell)\bigg( \nu_0^\ell + \sum_{j\neq\ell}\map{j}{\ell}(R^j) + \sum_{k=1}^{t-1}r_k^\ell \bigg)\Bigg),
\end{align}
where $S^j=\sum_{k=1}^{T^j}s_k^j$ and $R^j=\sum_{k=1}^{T^j}r_k^j$ were used in the last equality. Note also that the $\map{j}{\ell}$ operator returns zero for all statistics belonging to parameters in $\theta^j$ that are \textit{not} shared with dataset $\ell$, which enables incorporating statistics multiplied with $\theta^j$ and $A(\theta^j)$ in the first proportionality into the expressions multiplied with $\theta^\ell$ in the last equality.
Comparing \eqref{eq:Th_ellPost} with the parameter prior \eqref{eq:EfamPrior} it is clear that the posterior for $\theta^\ell$ has the same form as the prior, but with updated hyperparameters
\begin{align} 
&\chi_{t-1}^\ell = \chi^\ell_0 + \sum_{j\neq\ell} \map{j}{\ell}(S^j) + \sum_{k=1}^{t-1}s_k^\ell , \label{eq:ShyperParam1} \\
&\nu_{t-1}^\ell = \nu^\ell_0 + \sum_{j\neq\ell} \map{j}{\ell}(R^j) + \sum_{k=1}^{t-1}r_k^\ell \label{eq:ShyperParam2}.
\end{align}

Using the expressions for the updated hyperparameters in the parameter posterior, i.e.\ $p(\theta^\ell|x_{0:t-1}^\ell,y_{1:t-1}^\ell,\mathcal{D}^{\neg\ell})=p(\theta^\ell|\chi_{t-1}^\ell,\nu_{t-1}^\ell)$, we can evaluate \eqref{eq:SmargInt} analytically
\begin{align} \label{eq:SmargIntAnalytic}
p(x_t^\ell,y_t^\ell &| x_{0:t-1}^\ell,y_{1:t-1}^\ell,\mathcal{D}^{\neg \ell}) = \int p(x_t^\ell,y_t^\ell | x_{t-1}^\ell,\theta^\ell)p(\theta^\ell | \chi_t^\ell,\nu_t^\ell)\mathrm{d}\theta^\ell \nonumber \\
&= \int h_t^\ell \exp\left( {\theta^\ell}^\Tr s_t^\ell - {A^\ell}^\Tr(\theta^\ell)r_t^\ell \right)g^\ell(\chi_{t-1}^\ell,\nu_{t-1}^\ell) \exp\left( {\theta^\ell}^\Tr \chi_{t-1}^\ell - {A^\ell}^\Tr(\theta^\ell)\nu_{t-1}^\ell \right)  \mathrm{d}\theta^\ell \nonumber \\
&= h_t^\ell g^\ell(\chi_{t-1}^\ell,\nu_{t-1}^\ell) \int \exp\left( {\theta^\ell}^\Tr\left(\chi_{t-1}^\ell+s_t^\ell \right) - {A^\ell}^\Tr(\theta^\ell)\left( \nu_{t-1}^\ell + r_t^\ell \right) \right) \mathrm{d}\theta^\ell \nonumber \\
&= h_t^\ell \frac{g^\ell(\chi_{t-1}^\ell,\nu_{t-1}^\ell)}{g^\ell(\chi_{t}^\ell,\nu_{t}^\ell)} \int \underbrace{g^\ell(\chi_{t}^\ell,\nu_{t}^\ell) \exp\left( {\theta^\ell}^\Tr\chi_{t}^\ell - {A^\ell}^\Tr(\theta^\ell)\nu_{t}^\ell \right)}_{=p(\theta^\ell|\chi_t^\ell,\nu_t^\ell)} \mathrm{d}\theta^\ell = h_t^\ell \frac{g^\ell(\chi_{t-1}^\ell,\nu_{t-1}^\ell)}{g^\ell(\chi_{t}^\ell,\nu_{t}^\ell)},
\end{align}
where we used $\chi_t^\ell=\chi_{t-1}+s_t^\ell$ and $\nu_t^\ell=\nu_{t-1}+r_t^\ell$ in the second to last equality and the last equality is due to the parameter posterior integrating to one.

\section{Implementation Details: Linear-Gaussian Models} \label{app:impDetailLG}
This section briefly discusses some implementation details that are common for the two linear-Gaussian multi-SSMs used in the experiments in Section \ref{sec:multiSSM_mPG}. It also provides details on the initialization of the samplers for the linear-Gaussian multi-SSMs in Section \ref{sec:LGhier} and \ref{sec:NeuripsCompare}.

Marginalizing out a parameter with an inverse gamma prior distribution from a normally distributed likelihood yields a \textit{Students' t} distributed update. The normal distribution is \begin{equation}\label{eq:NormDist}
p(x|\mu,\theta) = \frac{1}{\sqrt{2\pi\theta}} \exp(-\frac{1}{2}\frac{(x-\mu)^2}{\theta}),
\end{equation}
where $\mu$ is the mean value and $\theta$ is the variance of the distribution, and its inverse gamma, conjugate prior is
\begin{equation}\label{eq:invGamDist}
p(\theta|\alpha,\beta) = \frac{\beta^\alpha}{\Gamma(\alpha)}\frac{1}{\theta^{\alpha+1}}\exp\left({-\frac{\beta}{\theta}}\right),
\end{equation}
where $\alpha$ and $\beta$ are hyperparameters. It can be shown that the hyperparameter updates for a normally distributed likelihood \eqref{eq:NormDist} with an inverse gamma conjugate prior \eqref{eq:invGamDist} are $\alpha_u = \alpha + \frac{1}{2}$ and $\beta_u = \beta + \frac{(x-\mu)^2}{2}$ and the marginal likelihood is
\begin{equation} \label{eq:StudentsT}
p(x|\nu,m,s) = \frac{\Gamma\left(\frac{2\alpha+1}{2}\right)}{\Gamma\left( \frac{2\alpha}{2} \right)\sqrt{\pi 2 \beta}}\left( 1+\frac{(x-\mu)^2}{2\beta} \right)^{-\frac{2\alpha+1}{2}}.
\end{equation}
The Student's t distribution is available in MATLAB, so draws from or evaluations of \eqref{eq:StudentsT} are easy to implement. The inverse gamma distribution is not implemented in MATLAB, so to draw samples from the inverse gamma posterior we instead draw samples from a gamma distributed random variable and compute the reciprocal. Note that the gamma distribution in MATLAB uses a different parameterization than that in \eqref{eq:invGamDist}, so the hyperparameters must be rescaled appropriately.

\subsection{Single marginalized PG for Linear-Gaussian multi-SSM}
To generate observation data, we simulated from \eqref{eq:LGSSM1}-\eqref{eq:LGSSM2} using parameter values $\varT_1=0.1$, $\varT_2=1$, $\varO_1=2$ and ${\varO_2=1}$. We used inverse gamma priors for all unknown noise variances, i.e.\ $\varT_1,\varT_2,\varO_1,\varO_2 \sim \mathcal{IG}(\alpha,\beta)$, with hyperparameters $\alpha=\beta=0.01$. 
The parameters were initialized to $\varT_1= 1$, $\varT_2=0.1$, $\varO_1=3$ and ${\varO_2=0.1}$.

\subsection{Linear-Gaussian model with shared process noise} \label{app:initialValLG}
We used the same observation and transition constants as for the linear\hyp{}Gaussian model in Section \ref{sec:LGhier}, i.e $a=0.9$ and $c=1$. Observation data was generated by simulating from \eqref{eq:LGstack1}-\eqref{eq:LGstack2} with parameter values $\varT=0.1$ and $\varO^\ell=1+2\cdot u$, where $u$ is a uniformly distributed random variable in the range $[0,1]$. The inverse gamma prior used hyperparameters $\alpha=1.5$ and $\beta=0.5$, and the parameters were initialized to $\varT=\varO^\ell= 5$. 

The integrated ACF was computed by summing the ACF values until the first lag where they crossed zero. If the ACF did not cross zero before the maximum lag considered, 100, the integrated ACF was set to the sum of the ACF up until the maximum lag. Unless the cSMC algorithm degenerated, the ACF crossed zero well before lag 100 in this example.

\section{Modeling the Spread of Mosquito-borne Diseases} \label{app:VBD}
This section gives a detailed description and motivation of the individual compartmental SSMs used to model the transmission of dengue and Zika virus in Section \ref{sec:VBD}. It also discusses the parameter priors that were used, and specifies the state initialization for all three outbreaks.

\subsection{The State Transition Model} \label{app:VBDtrans}
The SSM consists of two coupled compartmental models, one for the human population and one for the mosquito population.
The human population is divided into four compartments: susceptible, exposed, infectious and recovered. Once infected, the mosquitoes remain infectious for the rest of their lifespan, motivating using only three compartments for the mosquito population. The coupling between the human SEIR model and the mosquito SEI model is due to cross-infection between an infectious individual from one population and a susceptible individual from the other population. This model structure is visualized in Figure \ref{fig:VBDmodel} (right) in the main article.
The number of individuals in each compartment are the state variables, i.e.\ $x_t=(S_t^h, E_t^h, I_t^h, R_t^h, S_t^m, E_t^m, I_t^m)^\Tr$, where superscript $h$ indicates a variable for the human population and $m$ a variable for the mosquito population. The total population count is assumed to be constant, thus one of the state variables is in fact given by the remaining ones through the relations $n^h=S_t^h+E_t^h+I_t^h+R_t^h$ and $n^m=S_t^m+E_t^m+I_t^m$. The state transitions between compartments are stochastic and discrete in both state and time. They are given by the relations
\begin{equation}
\label{eq:SEIRstoch}
\begin{aligned} 
S^{h}_{t} &= S^{h}_{t-1} - e^{h}_{t},    			 & \hspace{2mm} \bar{S}^{m}_{t} &= S^{m}_{t-1} - e^{m}_{t},  \\
E^{h}_{t} &= E^{h}_{t-1} + e^{h}_{t} - i^{h}_{t},	 & \hspace{2mm} \bar{E}^{m}_{t} &= E^{m}_{t-1} + e^{m}_{t} - i^{m}_{t}, \\
I^{h}_{t} &= I^{h}_{t-1} + i^{h}_{t} - r^{h}_{t}, 	 & \hspace{2mm} \bar{I}^{m}_{t} &= I^{m}_{t-1} + i^{m}_{t},  \\
R^{h}_{t} &= R^{h}_{t-1} + r^{h}_{t}, 
\end{aligned}
\end{equation}
where lowercase letters indicate the stochastic number of newly exposed ($e_t$), infected ($i_t$) and recovered ($r_t$). For the mosquito population, the bar indicates the number in each compartment before taking births and deaths into account. The duration of the outbreaks is short enough for human births/deaths to be ignored.

The model for the stochastic transitions in \eqref{eq:SEIRstoch} is based on the discrete-state and discrete-time stochastic model by \cite{Murray2018}, which in turn is based on the continuous time and state deterministic model by \cite{Funk2016}. For the number of newly infected and recovered, we adopt the same binomial transition densities as in \cite{Murray2018}, i.e.\,
\begin{equation}
\label{eq:SEIR_ir_update}
\begin{aligned} 
i^{h}_{t} &\sim \text{Bin}(i^{h}_{t} \mid E^{h}_{t-1}, \delta^{h}),& \hspace{2mm}  i^{m}_{t} &\sim \text{Bin}(i^{m}_{t} \mid E^{m}_{t-1}, \delta^{m}),  \\
r^{h}_{t} &\sim \text{Bin}(r^{h}_{t} \mid I^{h}_{t-1}, \gamma^{h}),   &  &
\end{aligned}
\end{equation}
where $\delta$ is the infection probability, and $\gamma$ is the recovery probability. 

In \cite{Murray2018}, the number of newly exposed are drawn from binomial distributions in two steps that involve one parameter --- the disease dependent transmission probability, $\lambda$. In the first of the two updates a location dependent bite frequency, $b$, can be included (it is a constant, one bite/mosquito/day, in \cite{Murray2018}), but it occurs inside an exponential term in the success probability of the binomial distribution and cannot be marginalized out. We will therefore use an alternative model formulation to generate the number of newly exposed, where both the bite frequency and the transmission probability can be marginalized out.

The new susceptible-to-exposed update is derived for the human population. The derivation for the mosquito population is exactly analogue and those updates are given after the derivation for the human population.
In the new formulation, the total number of interactions, $K$, between mosquitoes and humans in each timestep is assumed to follow a Poisson distribution with the total number of bites during a day as rate parameter,
\begin{equation} \label{eq:VBBDTotalInteract}
K \sim \text{Po}\left(b n^m\right),
\end{equation}
where $b$ is the bite frequency in bites/mosquito/day. For each of the $K$ interactions, the probability that the interaction is between a susceptible human and an infectious mosquito is $\frac{S_{t-1}^h I_{t-1}^m}{n^h n^m}$. Aggregated over all $K$ interactions, the number of interactions where a susceptible human was potentially exposed to the virus, $K_\text{exp}^h$, can be drawn from
\begin{equation} \label{eq:Kexp}
K_\text{exp}^h \sim \text{Bin}\left( K, \frac{S_{t-1}^h I_{t-1}^m}{n^h n^m} \right).
\end{equation}
To avoid one sampling step, $K$ can be marginalized out from \eqref{eq:Kexp}, yielding the update
\begin{equation}
K_\text{exp}^h \sim \text{Po}\left( b\frac{S_{t-1}^h I_{t-1}^m}{n^h} \right).
\end{equation}
Each of the $K_{exp}^h$ interactions between a susceptible human and an infectious mosquito leads to infection with transmission probability $\lambda$. 
Aggregating over all $K_\text{exp}^h$ interactions, the number of infectious bites can be drawn from 
\begin{equation}
K_\text{inf}^h \sim \text{Bin}\left(K_\text{exp}^h,\lambda^h\right).
\end{equation}
Each susceptible human can (potentially) be bitten by more than one infectious mosquito, thus, the number of infectious bites must be distributed across the susceptible humans. 
This can be done with the \textit{classical occupancy distribution} (also called Stevens-Craig distribution), which models the number of non-zero counts in a multinomial distribution with $K_\text{inf}^h$ trials and equal event probabilities $1/S_{t-1}^h$ \citep{ONeill2021}. 
The number of newly exposed humans are drawn from
\begin{equation}
e_t^h \sim \text{Occ}(K_\text{inf}^h, S_{t-1}^h).
\end{equation}
The complete susceptible-to-exposed update for both populations are
\begin{equation} \label{eq:s2eUpdate}
\begin{aligned}
K_\text{exp}^h &\sim \text{Po}\left( b\frac{S_{t-1}^h I_{t-1}^m}{n^h} \right),& \hspace{2mm}   K_\text{exp}^m &\sim \text{Po}\left( b\frac{S_{t-1}^m I_{t-1}^h}{n^h} \right), \\
K_\text{inf}^h &\sim \text{Bin}\left(K_\text{exp}^h,\lambda^h\right),& \hspace{2mm}   K_\text{inf}^m &\sim \text{Bin}\left(K_\text{exp}^m,\lambda^m\right), \\
e_t^h &\sim \text{Occ}(K_\text{inf}^h, S_{t-1}^h),&  \hspace{2mm}   e_t^m &\sim \text{Occ}\left(K_\text{inf}^m, S_{t-1}^m\right).
\end{aligned}
\end{equation}
The complete parameter vector is $\theta=(\lambda^h,\delta^h,\gamma^h,\lambda^m,\delta^m,b)^\Tr$ for the state transition part of the model, where $b$ is location dependent and the remaining parameters are disease dependent. The state vector, where the number of susceptible has been excluded and the number of newly infected humans have been included since they contribute to the statistic for the reporting probability, is $x_t=(E_t^h, I_t^h, R_t^h, E_t^m, I_t^m, K_\text{exp}^h, K_\text{exp}^m, K_\text{inf}^h, K_\text{inf}^m, i_t^h)^\Tr$.

Another difference from the model by \cite{Murray2018} is that mosquito deaths and births are deterministic (rather than stochastic) updates of the state variables. Mosquitoes are born from parents in all compartments with a rate $\nu$, and mosquitoes in all compartments die with a rate $\mu$. Both of these rates are considered known and are set to  $\nu=\mu=1/7$ to reflect a life-span of one week \citep{Funk2016}. All new-born mosquitoes are assumed to be susceptible irrespective of their parents disease state. The final update of the mosquito states is
\begin{equation}
\begin{aligned}
S^{m}_{t} &= (1-\mu)\bar{S}_t^m + \nu n^m, \\
E^{m}_{t} &= (1-\mu)\bar{E}_t^m, \\
I^{m}_{t} &= (1-\mu)\bar{I}_t^m. \\
\end{aligned}
\end{equation}

\subsection{The Observation Model} \label{app:VBDobs}
The observations from the outbreaks are the number of newly infected humans each day (dengue fever) or each week (Zika) that visited a health center to get medical care. Since both diseases are known to often cause mild symptoms, the reported number of newly infected humans is likely to only correspond to a proportion of the true number of new infections. Therefore, the observations are assumed to follow the binomial distribution
\begin{equation} \label{eq:SEIRobs}
y_{t} \sim \text{Bin}\left(\sum_{j=t-l+1}^t i_{j}^h,\rho\right),
\end{equation}
where $\rho$ is the reporting probability. The newly infected humans are aggregated over the time interval between observations, $l$, which is one for dengue fever (daily observations) and seven for Zika (weekly observations). There is no indication that the reporting probability is predominantly disease- or location dependent, it is therefore assumed to be an outbreak-specific parameter.

\subsection{The Parameter Prior} \label{app:VBDparam}
The parameter priors are chosen to form suitable conjugacy relations with the corresponding transition or observation density to allow for marginalization. For all binomial updates, a suitable prior distribution for the unknown success probability is the beta distribution. 
The incubation time for humans and mosquitoes and the infection time for humans, are well-documented in the literature and can be used to design informed priors on the infection and recovery probabilities. The Gaussian priors on incubation and recovery rates used by \cite{Funk2016} can be transformed to beta priors on infection and recovery probabilities using mode-matching through the relation 
\begin{equation}\label{eq:betaMode}
\frac{\alpha-1}{\alpha+\beta-2} = \frac{1}{\mu_0},
\end{equation}
where $\alpha$ and $\beta$ are shape parameters of the beta distribution, and $\mu_0$ is the mean transition time for the Gaussian distribution. One possible choice, used by \cite{Murray2018}, is $\alpha=1+\frac{2}{\mu0}$ and $\beta=3-\frac{2}{\mu0}$. Incubation and infection times are similar for Zika and dengue fever, therefore the same hyperparameters $\alpha$ and $\beta$ are used for both \citep{Funk2016}.
No prior information is available for the transmission probabilities $\lambda$ and the reporting probabilities $\rho$, and they are therefore assigned close to uniform  beta priors ($\alpha=\beta=1.05$ or $\alpha=\beta=1.1$). 

To establish a conjugacy relation between the bite frequency, $b$, and the number of interactions between susceptible and infectious individuals, $K_\text{exp}^h$ and $K_\text{exp}^m$, we will adopt a gamma prior for the bite frequency. It should be noted that because the bite frequency is present in two updates in \eqref{eq:s2eUpdate}, its posterior will depend on both $K_\text{exp}^h$ and $K_\text{exp}^m$. The bite frequency of mosquitoes on Yap and Fais is not well-studied and is intimately related to the unknown size of the mosquito population in this model. The earlier study by \cite{Funk2016} observed that the outbreak develops much quicker on the smaller island Fais, which motivates using a prior with a larger mean for the bite frequency on Fais. Furthermore, mosquitoes have been known to bite more than once per day, motivating the use of a prior that captures a range of bite frequencies, both larger and smaller than one \citep{Harrington2014}. Based on the above, the gamma prior hyperparameters for the bite frequency were set to $\alpha=1.2$, $\beta=1$ on Yap and $\alpha=3$, $\beta=1$ on Fais.

To summarize, the disease-specific parameter priors are
\begin{equation} \label{eq:SEIRprior}
\begin{aligned}
\lambda^h &\sim \text{Beta}(1.05,1.05),& \hspace{4mm} \lambda^m &\sim \text{Beta}(1.1,1.1), \\
\delta^h  &\sim \text{Beta}(1+\frac{2}{4.4},3-\frac{2}{4.4}),& \hspace{4mm} \delta^m  &\sim \text{Beta}(1+\frac{2}{6.5},3-\frac{2}{6.5}),  \\
\gamma^h  &\sim \text{Beta}(1+\frac{2}{4.5},3-\frac{2}{4.5}),& & \\
\end{aligned}
\end{equation}
for both dengue and Zika outbreaks (10 parameters in total). The location-specific parameter priors are
\begin{equation}
b^Y \sim \text{Gam}\left(1.2,1\right) ,  \hspace{4mm} 	b^F \sim \text{Gam} \left(3,1\right), 
\end{equation}
and the outbreak-specific parameter priors, one for each outbreak, are
\begin{equation}
\rho \sim \text{Beta}(1.05,1.05).
\end{equation}

\subsection{Initial State Values} \label{app:VBDinit}
The human population size, $n^h$, is 7370 on Yap and 294 on Fais. The outbreaks are assumed to start in the human population, so the initial number of exposed and infectious mosquitoes are set to zero for all three outbreaks. 
The initial number of exposed and infected humans are generated from Poisson distributions with individual rate parameters for each outbreak chosen based on the population size, the final outbreak size and how far the outbreak has progressed at the initial observation. The initial number of infected humans includes a ``-1'' to ensure that there is always at least one infected human when the simulation starts.
There have been previous outbreaks of dengue fever on Yap implying that the population there have some immunity. To reflect the unknown level of immunity in the Yap population, the initial number of recovered humans is drawn uniformly from $[0,n^h-E^h_0-I^h_0]$. There have not been any known earlier outbreaks of Zika on Yap or dengue on Fais, which motivates setting the initial number of recovered humans for these outbreaks to zero.

The number of mosquitoes on Yap and Fais that interact with the human population is not known apriori. Since the size of the mosquito population and the bite frequency together determines the number of interactions between mosquitoes and humans, they will be difficult to identify independently. We therefore keep the mosquito population fixed to $n^m=c\cdot n^h$ and let the bite frequency vary. The parameter $c$ is a scaling factor. In our simulations we have used $c=10$ for all outbreaks.

The initial state values for the outbreak of dengue fever on Yap are
\begin{equation} \label{eq:initialsYap}
\begin{aligned}
E^{h}_{0}  &\sim \text{Poisson}(15),  & E^{m}_{0}&=0,  \\
I^{h}_{0}-1&\sim \text{Poisson}(15),  & I^{m}_{0}&=0,  \\
R^{h}_{0}  &\sim \text{Uniform}(0,n^h-E^{h}_{0}-I^{h}_{0}), 	 &S^{m}_{0}&=c\cdot n^h,   \\
S^{h}_{0}  &=n^h-E^{h}_{0}- I^{h}_{0}- R^{h}_{0}.	 & &
\end{aligned}
\end{equation}
The initial state values for the outbreak of dengue fever on Fais are
\begin{equation} \label{eq:initialsFais}
\begin{aligned}
E^{h}_{0}  &\sim \text{Poisson}(2),  & E^{m}_{0}&=0,  \\
I^{h}_{0}-1&\sim \text{Poisson}(2),  & I^{m}_{0}&=0,  \\
R^{h}_{0}  &=0, 	 &S^{m}_{0}&=c\cdot n^h,   \\
S^{h}_{0}  &=n^h-E^{h}_{0}- I^{h}_{0}- R^{h}_{0}.	 & &
\end{aligned}
\end{equation}
The initial state values for the outbreak of Zika on Yap are
\begin{equation} \label{eq:initialsZika}
\begin{aligned}
E^{h}_{0}  &\sim \text{Poisson}(12),  & E^{m}_{0}&=0,  \\
I^{h}_{0}-1&\sim \text{Poisson}(12),  & I^{m}_{0}&=0,  \\
R^{h}_{0}  &=0, 	 &S^{m}_{0}&=c\cdot n^h,   \\
S^{h}_{0}  &=n^h-E^{h}_{0}- I^{h}_{0}- R^{h}_{0}.	 & &
\end{aligned}
\end{equation}

\subsection{Implementation Details} \label{app:VBDimplement}
States are propagated on the timescale of days for all three outbreaks to make updates of coupled parameters simpler and to ensure that an individual has the possibility to move from susceptible to exposed to infectious within a week (not possible with weekly state updates). Weight updates are performed daily for the dengue datasets and weekly for the Zika outbreak.

Sampling from the binomial distribution, $\text{Bin}(n,\theta)$ with number of trials $n$ and success probability $\theta$, is computationally expensive for large population sizes, but can be approximated by a normal distribution with mean $n\theta$ and variance $n\theta(1-\theta)$. The normal approximation is used whenever the two criteria
\begin{equation}
n>k\left(\frac{1-\theta}{\theta}\right), \quad n>k\left(\frac{\theta}{1-\theta}\right),
\end{equation}
are fulfilled. The parameter $k$ controls the accuracy of the approximation. We use $k=100$, which implies approximating only when it can be guaranteed that the range of possible values for the binomial distribution, $(0,n)$, are within 10 standard deviations of the mean of the normal approximation.

When a parameter with a beta prior distribution is marginalized out from a binomial update the resulting update is \textit{betabinomial}. To see this, note that the binomial distribution is
\begin{equation}\label{eq:BinDist}
p(x|n,\theta) = \binom{n}{x}\theta^x(1-\theta)^{n-x},
\end{equation}
and its beta conjugate prior is
\begin{equation}\label{eq:BetaDist}
p(\theta|\alpha,\beta)=\frac{\Gamma(\alpha+\beta)}{\Gamma(\alpha)\Gamma(\beta)}\theta^{\alpha-1}(1-\theta)^{\beta-1},
\end{equation}
where $\alpha$ and $\beta$ are hyperparameters. It can be shown that for the binomial distribution \eqref{eq:BinDist} with parameter prior \eqref{eq:BetaDist} the hyperparameter updates are $\alpha_u = \alpha + x$ and $\beta_u = \beta + n - x$ and the marginal likelihood is 
\begin{equation} \label{eq:BetaBinML}
p(x|\alpha,\beta) =\binom{n}{x}\frac{B(\alpha+x,\beta+n-x)}{B(\alpha,\beta)}.
\end{equation}
There is no implementation of the beta-binomial distribution in MATLAB. Therefore, sampling from this distribution is done by first drawing a sample from the beta posterior distribution (with hyperparameters updated based on earlier updates) and then using that sample in the binomial distribution. If applicable, the normal approximation described in the preceding paragraph is used to alleviate the computational burden of sampling from the binomial distribution for large population sizes. In addition MATLAB's built-in binomial sampler \textit{binornd} is exceptionally slow for large populations, so instead we use the user-implemented binomial sampler \textit{bnldev}\footnote{Available at \url{https://www.mathworks.com/matlabcentral/fileexchange/42464-binomial-random-number-generator}}.

Marginalizing out a parameter with a gamma prior distribution from a Poisson update results in a \textit{negative binomial} update. To see this, note that the Poisson distribution is
\begin{equation}\label{eq:PoiDist}
p(x|\theta) = \frac{\theta^x \exp(-\theta)}{x!},
\end{equation}
where $\theta$ is a rate parameter, and its gamma conjugate prior is
\begin{equation} \label{eq:GamDist}
p(\theta|\alpha,\beta) = \frac{\beta^\alpha}{\Gamma(\alpha)}\theta^{\alpha-1}\exp(-\beta\theta),
\end{equation}
where $\alpha$ and $\beta$ are hyperparameters. It can be shown that for the Poisson distribution \eqref{eq:PoiDist} with parameter prior \eqref{eq:GamDist} the hyperparameter updates are $\alpha_u = \alpha + x$ and $\beta_u = \beta +1$ and the marginal likelihood is 
\begin{equation} \label{eq:PoiGamML}
p(x|\alpha,\beta) = \frac{\Gamma(\alpha+x)}{\Gamma(\alpha)x!}\left(\frac{\beta}{\beta+1}\right)^{\alpha}\left(\frac{1}{\beta+1}\right)^{x}.
\end{equation}
The negative binomial distribution is available in MATLAB, so this update is straight-forward to implement. 

Sampling from the \textit{classical occupancy distribution} can be implemented by first drawing the number of bites, $z_i$, allocated to susceptible individual $i$ from the multinomial distribution
\begin{equation}
(z_1,\dots,z_{s_{t-1}^h}) \sim \text{multi}(K_\text{inf}^h,1/s_{t-1}^h),
\end{equation}
and then summing over all individuals $i$ that received at least one bite
\begin{equation}
e_t^h = \sum_{i=1}^{s_{t-1}^h}\mathbb{1}(z_i>0).
\end{equation}

\section{Additional Results} \label{app:results}
This section contains additional results for the experiments presented in the article.

\subsection{Additional Results for the Linear-Gaussian Multi-SSM With Three Datasets and Four Shared Parameters} \label{app:Additional_LG}
\begin{figure}[H]
	\centering
	\includegraphics[width=.65\textwidth]{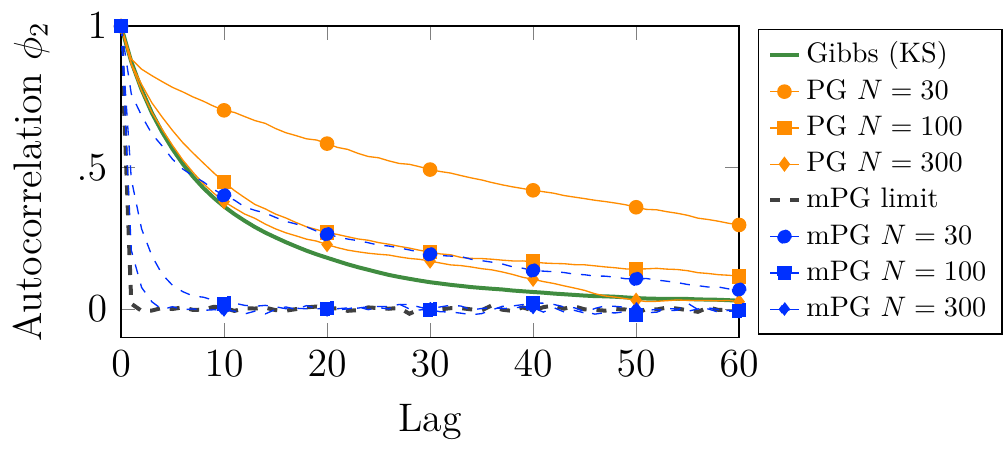}
	\caption{The autocorrelation function (ACF) for the parameter $\varT_2$ in Figure \ref{fig:LGhierSSM}. The ACF for PG converges to that of a Gibbs sampler with a Kalman smoother as $N$ increases, whereas the single marginalized PG sampler converges to a different, but lower limiting ACF as $N$ increases.}
	\label{fig:acfLG3_Q2}
\end{figure}

\begin{figure}[H]
	\centering
	\includegraphics[width=.65\textwidth]{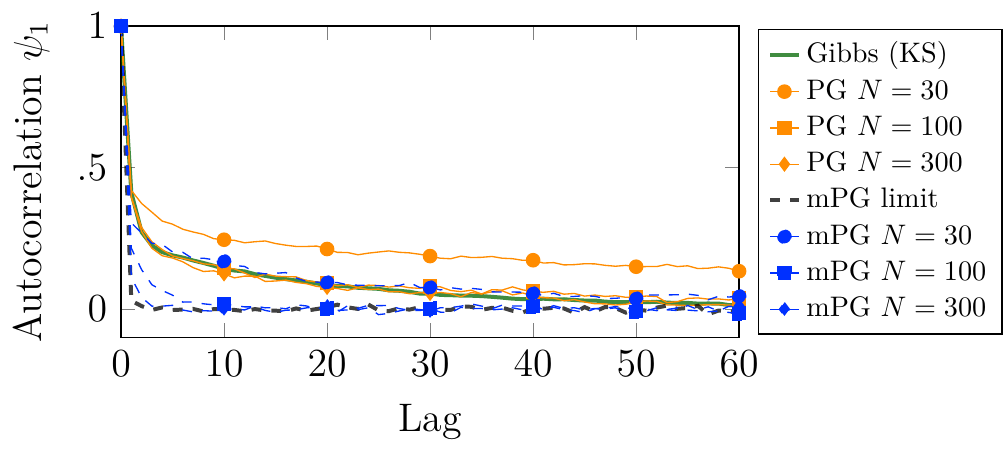}
	\caption{The autocorrelation function (ACF) for the parameter $\varO_1$ in Figure \ref{fig:LGhierSSM}. The ACF for PG converges to that of a Gibbs sampler with a Kalman smoother as $N$ increases, whereas the single marginalized PG sampler converges to a different, but lower limiting ACF as $N$ increases. }
	\label{fig:acfLG3_R1}
\end{figure}

\begin{figure}[H]
	\centering
	\includegraphics[width=.65\textwidth]{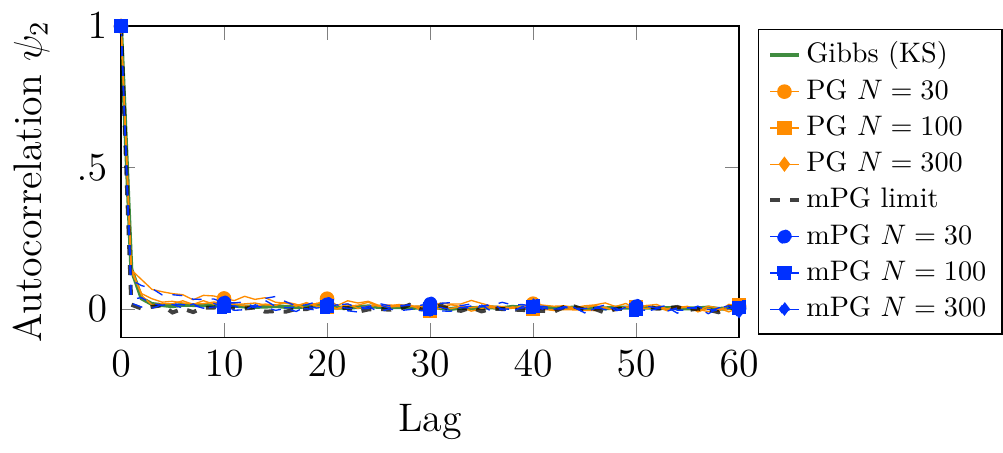}
	\caption{The autocorrelation function (ACF) for the parameter $\varO_2$ in Figure \ref{fig:LGhierSSM}. The ACF for PG converges to that of a Gibbs sampler with a Kalman smoother as $N$ increases, whereas the single marginalized PG sampler converges to a different, but lower limiting ACF as $N$ increases. }
	\label{fig:acfLG3_R2}
\end{figure}

Figure \ref{fig:acfLG3_Q2}, \ref{fig:acfLG3_R1} and \ref{fig:acfLG3_R2} show the ACFs for both the standard PG sampler and the single marginalized PG sampler for the parameters $\varT_2$, $\varO_1$ and $\varO_2$, respectively. For all three parameters the ACF of the standard PG sampler converges to the ACF of the Gibbs sampler as the number of particles increases, while the ACF of the single marginalized PG sampler converges to a limiting ACF below that of the Gibbs sampler.

\begin{figure}[H]
	\centering
	\includegraphics[width=.45\textwidth]{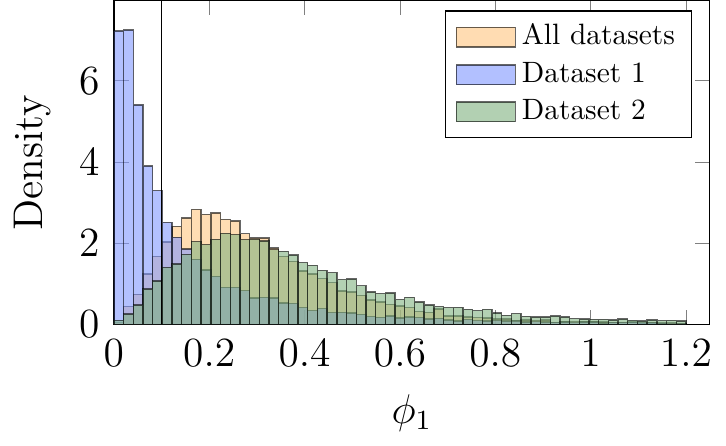}
	\caption{The histogram for the parameter $\varT_1$ in Figure \ref{fig:LGhierSSM} obtained from running a single marginalized PG sampler, both for the multi-SSM that uses all datasets for parameter estimation and for the case when all datasets are treated separately.}
	\label{fig:Histogram_3datasets_Q1}
\end{figure}

\begin{figure}[H]
	\centering
	\includegraphics[width=.45\textwidth]{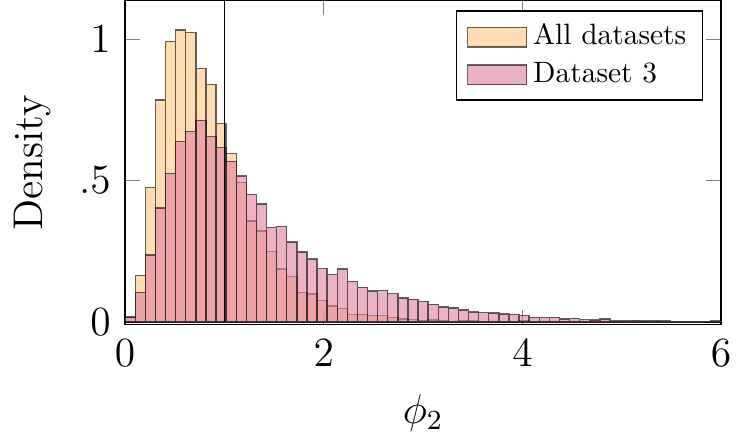}
	\caption{The histogram for the parameter $\varT_2$ in Figure \ref{fig:LGhierSSM} obtained from running a single marginalized PG sampler, both for the multi-SSM that uses all datasets for parameter estimation and for the case when all datasets are treated separately.}
	\label{fig:Histogram_3datasets_Q2}
\end{figure}

\begin{figure}[H]
	\centering
	\includegraphics[width=.45\textwidth]{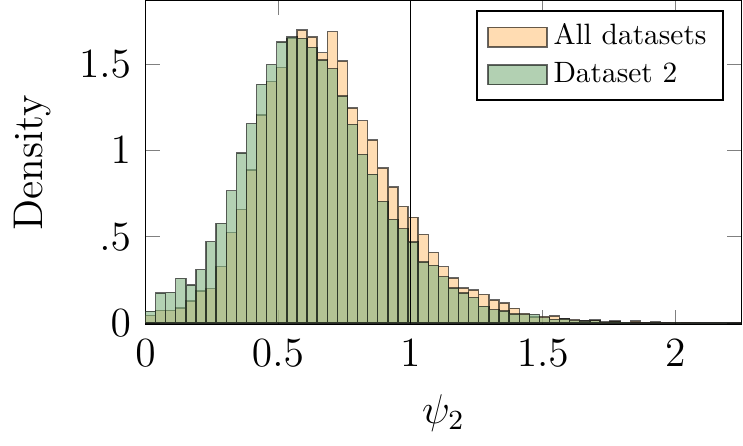}
	\caption{The histogram for the parameter $\varO_2$ in Figure \ref{fig:LGhierSSM} obtained from running a single marginalized PG sampler, both for the multi-SSM that uses all datasets for parameter estimation and for the case when all datasets are treated separately.}
	\label{fig:Histogram_3datasets_R2}
\end{figure}

Figure \ref{fig:Histogram_3datasets_Q1}, \ref{fig:Histogram_3datasets_Q2} and \ref{fig:Histogram_3datasets_R2} show the posterior distribution of the parameters $\varT_1$, $\varT_2$ and $\varO_2$, respectively, when using all three datasets as opposed to using just one of the datasets. The single marginalized PG sampler was used for estimation.

\subsection{Additional Results for the Comparison Between Single and Stacked Marginalized Particle Gibbs} \label{app:Additional_LGstack}
\begin{figure}[h]
	\centering
	\includegraphics[width=.45\textwidth]{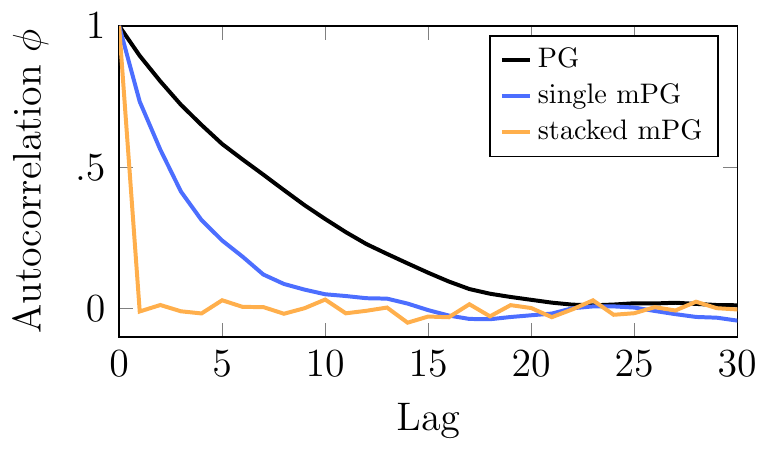}
	\caption{The autocorrelation function (ACF) for the process noise parameter $\varT$ shared among all four datasets. The ACF of the stacked marginalized PG sampler shows that it generates (close to) iid samples whereas the ACF for PG corresponds to that of an underlying (and correlated) Gibbs sampler and the ACF for the single marginalized PG sampler yields samples that are not iid, but have a weaker correlation than those from the PG sampler.
	}
	\label{fig:acf_L=4_Q}
\end{figure}

\begin{figure}[h]
	\centering
	\includegraphics[width=.8\textwidth]{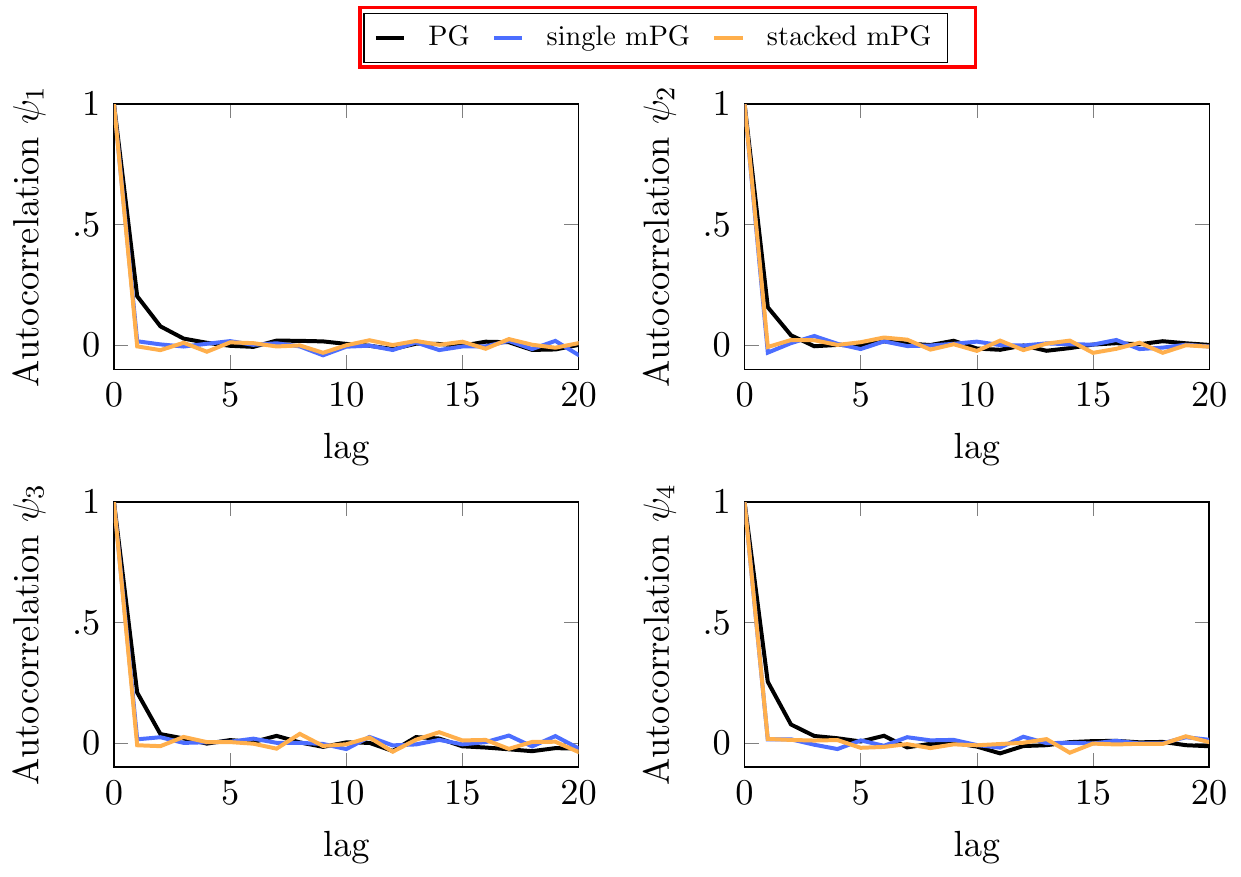}
	\caption{The autocorrelation function for the four observation noise parameters $\varO_1$, $\varO_2$, $\varO_3$ and $\varO_4$. All samplers yield only weakly correlated samples.}
	\label{fig:acf_L=4_R}
\end{figure}

The stacked marginalized PG sampler updates all state trajectories simultaneously and can therefore  generate (close to) iid samples. We illustrate this for the case of four datasets that share process noise parameters, but have individual observation noise parameters. All model parameters are the same as in the example in Section \ref{sec:NeuripsCompare}, but we use $N=100000$ particles (instead of $N=100$) for all three samplers to visualize the limiting ACFs of each sampler. Figure \ref{fig:acf_L=4_Q} shows the ACF for the shared transition noise parameter $\varT$ for PG, single marginalized PG and stacked marginalized PG.
It is clear that the stacked marginalized PG sampler indeed yields (close to) idd samples, whereas the single marginalized sampler yields samples with a weaker correlation than the PG sampler but they are not iid.
For completeness we have also included the corresponding ACFs for the four observation noise parameters $\varO_1$, $\varO_2$, $\varO_3$ and $\varO_4$ in Figure \ref{fig:acf_L=4_R}. In this case the two marginalized samplers perform similarly and even the PG sampler yields only weakly correlated samples.

\subsection{Additional Results for the Mosquito-borne Disease Multi-SSM} \label{app:Additional_VBD}
\begin{figure}[H]
	\centering
	\includegraphics[width=.75\textwidth]{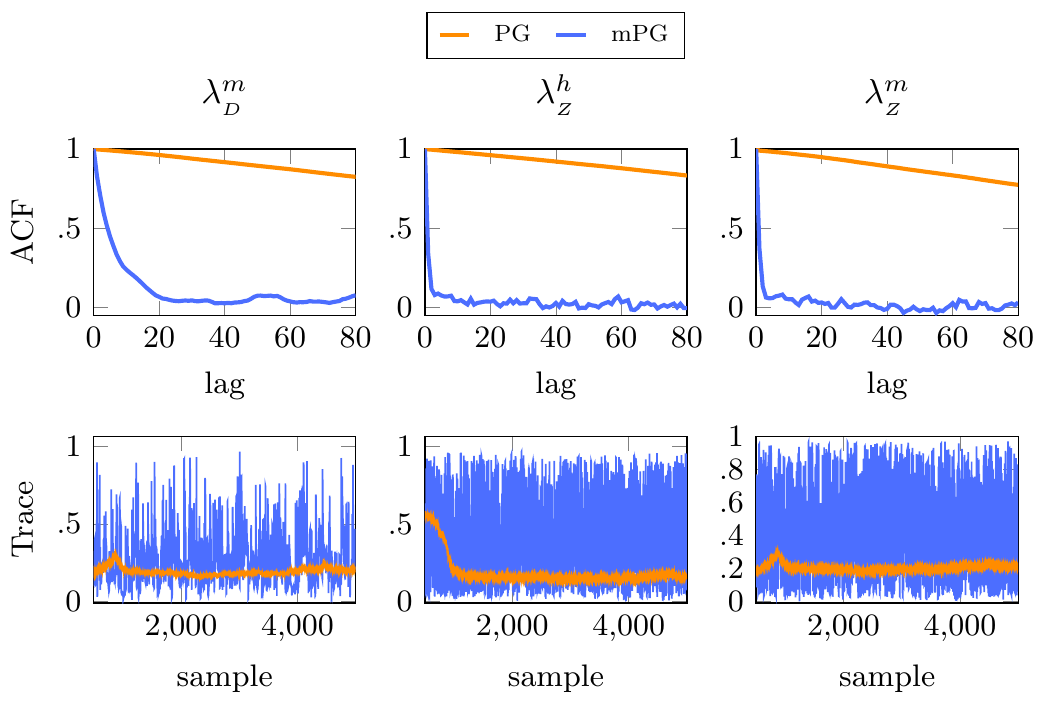}
	\caption{ACF and trace plot for the disease dependent parameters $\lambda^m_{_D}$, $\lambda^h_{_Z}$ and $\lambda^m_{_Z}$. While the standard PG sampler experiences poor mixing, the marginalized PG sampler \eqref{eq:kappaDenE}-\eqref{eq:kappaYapO} mixes well for all parameters. The corresponding histograms for the marginalized PG sampler are visualized in Figure \ref{fig:Histogram_VBD_1vs3_supp}, where the PG sampler has been excluded due to its poor mixing.}
	\label{fig:VBD_PGmPG_lambdas}
\end{figure}

\begin{figure}[H]
	\centering
	\includegraphics[width=.75\textwidth]{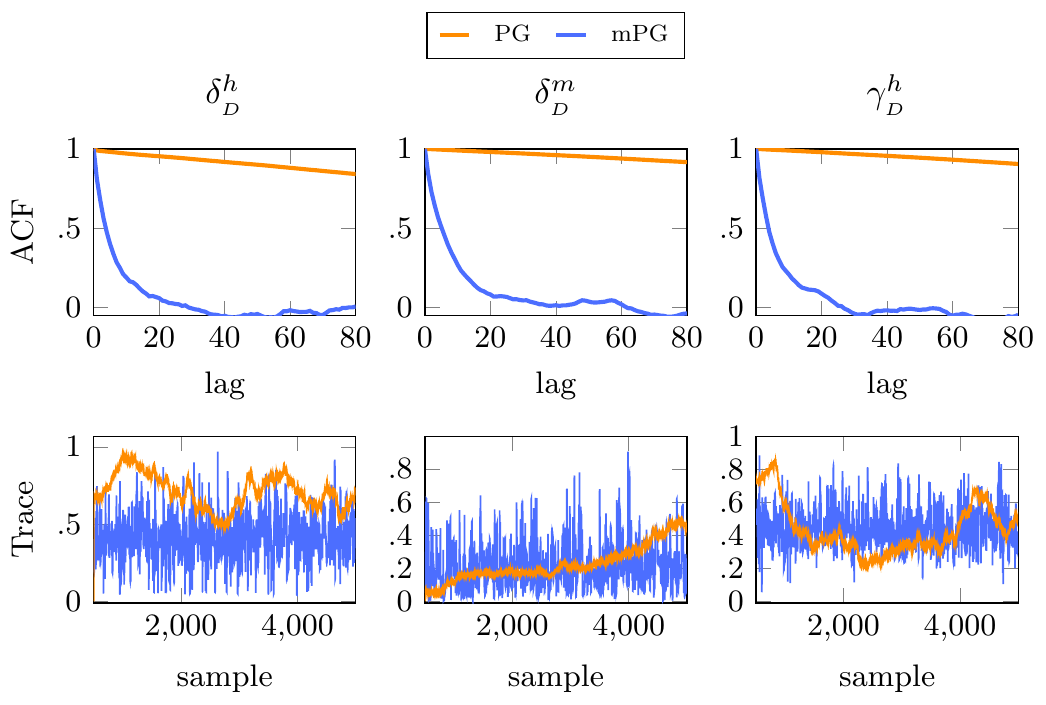}
	\caption{ACF and trace plot for the disease dependent parameters $\delta^h_{_D}$, $\delta^m_{_D}$ and $\gamma^h_{_D}$. While the standard PG sampler experiences poor mixing, the marginalized PG sampler \eqref{eq:kappaDenE}-\eqref{eq:kappaYapO} mixes well for all parameters. The corresponding histograms for the marginalized PG sampler are visualized in Figure \ref{fig:Histogram_VBD_1vs3_supp}, where the PG sampler has been excluded due to its poor mixing.}
	\label{fig:VBD_PGmPG_disDen}
\end{figure}

\begin{figure}[H]
	\centering
	\includegraphics[width=.75\textwidth]{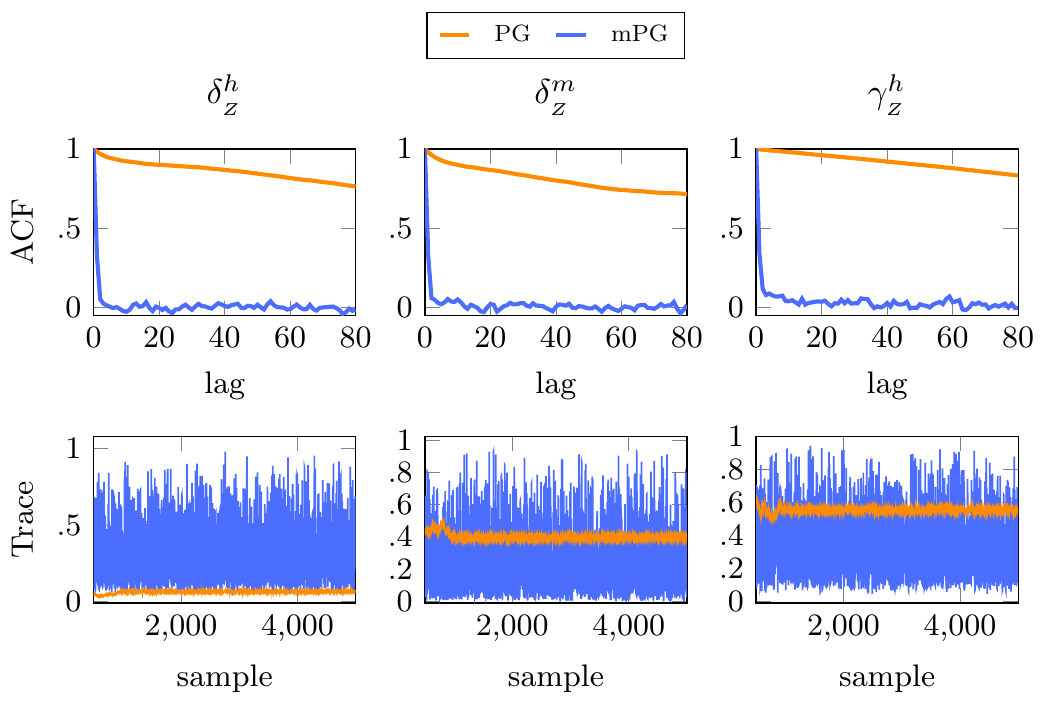}
	\caption{ACF and trace plot for the disease dependent parameters $\delta^h_{_Z}$, $\delta^m_{_Z}$ and $\gamma^h_{_Z}$. While the standard PG sampler experiences poor mixing, the marginalized PG sampler \eqref{eq:kappaDenE}-\eqref{eq:kappaYapO} mixes well for all parameters. The corresponding histograms for the marginalized PG sampler are visualized in Figure \ref{fig:Histogram_VBD_1vs3_supp}, where the PG sampler has been excluded due to its poor mixing.}
	\label{fig:VBD_PGmPG_disZika}
\end{figure}

\begin{figure}[H]
	\centering
	\includegraphics[width=.75\textwidth]{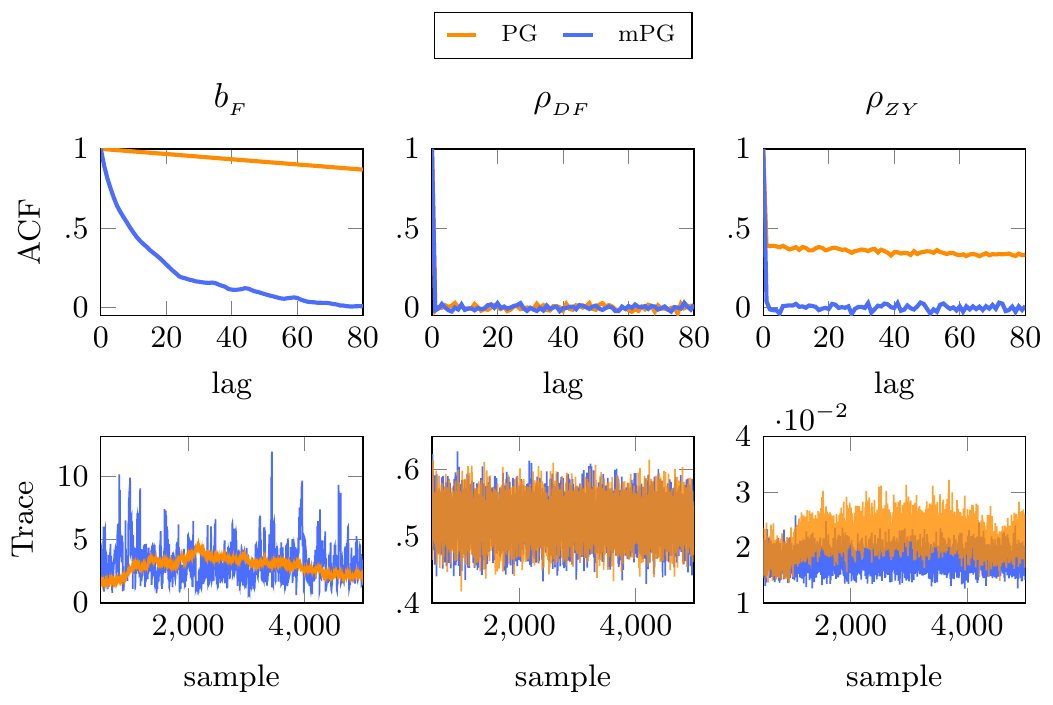}
	\caption{ACF and trace plot for the location dependent parameter $b_{_F}$ and the outbreak dependent parameters $\rho_{_{DF}}$ and $\rho_{_{ZY}}$. For better visualization, the trace plots for $\rho_{_{DF}}$ and $\rho_{_{ZY}}$ do not range between zero and one. While the standard PG sampler experiences poor mixing, the marginalized PG sampler \eqref{eq:kappaDenE}-\eqref{eq:kappaYapO} mixes well for all three parameters. The corresponding histograms for the marginalized PG sampler are visualized in Figure \ref{fig:Histogram_VBD_1vs3_supp}, where the PG sampler has been excluded due to its poor mixing.}
	\label{fig:VBD_PGmPG_brho}
\end{figure}

Figure \ref{fig:VBD_PGmPG_lambdas}, \ref{fig:VBD_PGmPG_disDen}, \ref{fig:VBD_PGmPG_disZika} and \ref{fig:VBD_PGmPG_brho} illustrate the ACF and trace plots for the disease dependent parameters $\lambda^m_{_D}$, $\lambda^h_{_Z}$, $\lambda^m_{_Z}$, $\delta^h_{_D}$, $\delta^m_{_D}$, $\gamma^h_{_D}$, $\delta^h_{_Z}$, $\delta^m_{_Z}$ and $\gamma^h_{_Z}$; the location dependent parameter $b_{_Y}$; and the disease dependent parameters $\rho_{_{DF}}$ and $\rho_{_{ZY}}$. In accordance with Figure \ref{fig:PGvsmPG_VBD} in the main article, the PG sampler tends to get stuck and produces highly correlated samples. Concurrently, the marginalized PG sampler explores the parameter space much more efficiently and yields only weakly correlated samples. 

\begin{figure}[H]
	\centering
	\includegraphics[width=\textwidth]{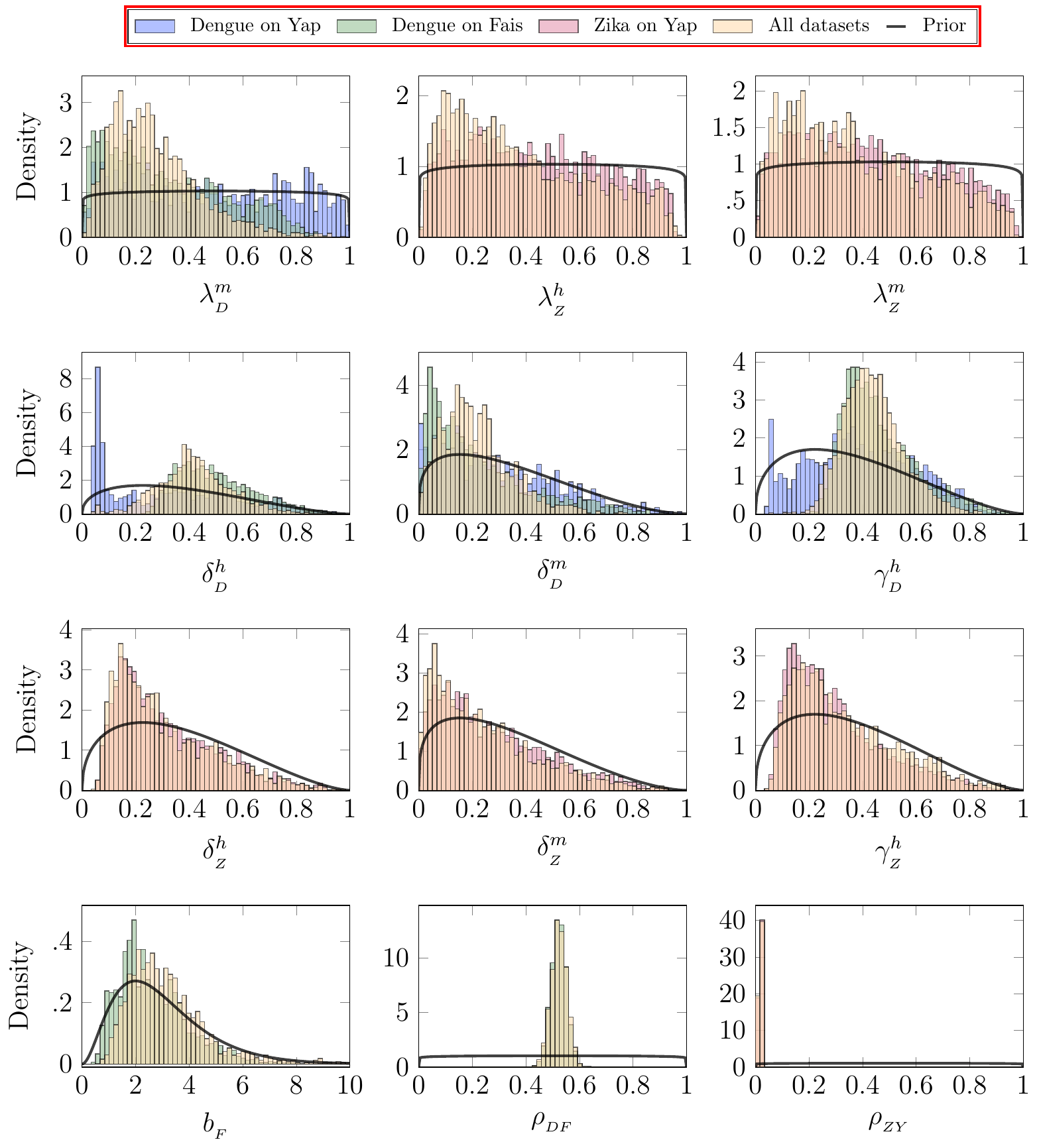}
	\caption{The histograms for all parameters except $\lambda^h_{_D}$, $b_{_Y}$ and $\rho^h_{_{DY}}$, which were presented in the main article, obtained from running the marginalized PG sampler \eqref{eq:kappaDenE}-\eqref{eq:kappaYapO}, both for the multi-SSM that uses all datasets for parameter estimation and for the case when all datasets are treated separately.}
	\label{fig:Histogram_VBD_1vs3_supp}
\end{figure}

Figure \ref{fig:Histogram_VBD_1vs3_supp} compares the sample-based parameter posterior obtained when using a single dataset with that obtained using the multi-SSM formulation for the disease dependent parameters $\lambda^m_{_D}$, $\lambda^h_{_Z}$, $\lambda^m_{_Z}$, $\delta^h_{_D}$, $\delta^m_{_D}$, $\gamma^h_{_D}$, $\delta^h_{_Z}$, $\delta^m_{_Z}$ and $\gamma^h_{_Z}$; the location dependent parameter $b_{_Y}$; and the disease dependent parameters $\rho_{_{DF}}$ and $\rho_{_{ZY}}$. These results indicate that the multi-SSM formulation is useful in the sense that the posterior is, for many of the parameters, less similar to the prior and more concentrated --- likely due to the aggregated information from all datasets.

\end{document}